\documentclass[twocolumn,showpacs,amsmath,amssymb,prd]{revtex4}
\def\ba{\begin{eqnarray}}
\def\ea{\end{eqnarray}}

\usepackage{graphicx}% Include figure files
\usepackage{dcolumn}% Align table columns on decimal point
\usepackage{bm}% bold math
\usepackage{epsf}

\begin{document}
%\twocolumn[\hsize\textwidth\columnwidth\hsize\csname@twocolumnfalse\endcsname

\title{Relativistic Hydrodynamic Evolutions with Black Hole Excision}

\author{Matthew D. Duez${}^1$, Stuart L. Shapiro${}^{1,2}$,
  and Hwei-Jang Yo${}^{1,3}$
 }

\address{
${}^1$ Department of Physics, University of Illinois 
 	at Urbana-Champaign, Urbana, IL 61801 \\
${}^2$ Department of Astronomy, \& NCSA, 
	University of Illinois at Urbana-Champaign, Urbana, IL 61801 \\
${}^3$ Institute of Astronomy and Astrophysics, Academia Sinica,
	Taipei 115, Taiwan, Republic of China }

\begin{abstract}
We present a numerical code designed to study astrophysical
phenomena involving dynamical spacetimes containing black holes
in the presence of relativistic hydrodynamic matter.  We present
evolutions of the collapse of a fluid
star from the onset of collapse to the settling of the resulting
black hole to a final stationary state.  In order to evolve stably
after the black hole forms, we
excise a region inside the hole before a singularity is encountered.  This
excision region is introduced after the appearance of an apparent horizon, but
while a significant amount of matter remains outside the hole.  We
test our code by evolving accurately a vacuum Schwarzschild black hole,
a relativistic Bondi accretion flow onto a black hole, Oppenheimer-Snyder
dust collapse,
and the collapse of nonrotating and rotating stars.  These systems are
tracked reliably for hundreds of $M$ following excision,
where $M$ is the mass of the black hole.  We perform these tests
both in axisymmetry
and in full 3+1 dimensions.  We then apply our code to study the
effect of the stellar spin
parameter $J/M^2$ on the final outcome of gravitational collapse of
rapidly rotating $n = 1$ polytropes.  We find that a black hole forms only if
$J/M^2<1$, in agreement with previous simulations.  When $J/M^2>1$, the collapsing
star forms a torus which fragments into nonaxisymmetric clumps, capable
of generating appreciable ``splash'' gravitational radiation.

\end{abstract}

%\pacs{PACS numbers:  04.30.Db, 04.25.Dm, 97.80.Fk}

\maketitle

%\draft

%\vskip2pc]

\section{Introduction}

Since many of the most interesting phenomena in astrophysics
involve black holes, the modeling of black hole spacetimes is
one the most important problems in numerical general relativity. 
It is also one of the most challenging problems.  Black hole evolutions
present all the usual difficulties of numerical relativity, such as
the need to find a stable form of the field evolution equations and
the need to find a practical coordinate system.  In addition, handling
the singular region is very subtle for a numerical code; the black hole
singularity must be avoided to allow the exterior evolution to continue
far into the future.

One of the most promising methods to date of dealing with black hole
singularities is black hole excision.  This method, first suggested
by Unruh~\cite{u84}, exploits the fact that the singularity resides
inside an event horizon,
a region that is casually disconnected from the rest of the
universe.  Since no physical information propagates from inside the
event horizon to outside, one should be able to evolve the exterior
independent of the interior spacetime.  Inside the event horizon,
causality entitles us to do
anything which will produce a stable exterior evolution.  In particular, one
can excise a region inside the horizon containing the singularity
and replace it with suitable boundary conditions at its outer surface.

Although it is guaranteed that no physical signal can propagate from
inside the horizon to outside, unphysical signals often can propagate
in evolution codes.  Gauge
modes can move acausally for many gauge conditions.  Although they
carry no physical content, such modes may destabilize the code.  Thus,
the choice of gauge is crucial to obtaining good excision evolutions. 
In addition, constraint-violating modes can, for some formulations of
the field equations, propagate acausally, creating inaccuracies and
instabilities.  Thus, the choice of formulation is also crucial to
obtaining good excision evolutions.

The feasibility of black hole excision was demonstrated in
spherically symmetric 1+1 dimensional evolutions of a single black
hole in the presence of a self-gravitating scalar field
\cite{ss92,mc96,admss95,gmw97}.
%, first by
%Seidel and Suen~\cite{ss92}, and then by Marsa and Choptuik~\cite{mc96},
%by Anninos {\it et al}~\cite{admss95}, and by
%Gomez {\it et al}~\cite{gmw97}.
%Meanwhile, Scheel {\it et al.}
Excision was also implemented successfully to study the spherically
symmetric collapse of collisionless matter to a black hole in Brans-Dicke
theory~\cite{sst95}.
Three-dimensional evolutions of black holes with excision were
attempted by using the standard 3+1 ADM formulation, for
a stationary~\cite{amsst95} and for a boosted black hole~\cite{BBHGCA98}.
Although the introduction of excision improved the behavior of these black
hole simulations, long-term stability could not be achieved due to
instabilities endemic to the unmodified ADM formulation.

Since then, new and more stable formulations of the 3+1 Einstein field
equations have been devised.  Using excision in a modified version of the
ADM equations commonly referred to as BSSN~\cite{sn95,bs99}, several groups
have evolved stationary black hole spacetimes (non-spinning and spinning)
for arbitrarily long times~\cite{ab01,ybs02}.  Long-term
stability has also been achieved using hyperbolic formulations of
the field equations~\cite{sklpt02,clnprst03,tln03} and using characteristic
evolutions~\cite{glmw98}.
Success has also been achieved in evolving distorted and moving
black holes with excision, both with characteristic formalisms~\cite{glmw98}
and with BSSN~\cite{abpst01,ssslsf03,sskls03}.  Excision has also been
used to simulate the grazing collision of two black holes~\cite{betal00}
and to simulate binary black holes for approximately one orbital
period~\cite{btj03}.

The last several years also have seen significant advances in
numerical, 3+1 relativistic hydrodynamics in dynamical spacetimes
(see, e.g.~\cite{s99,fgimrssst02,dmsb03}). 
The simulation of rapidly rotating, relativistic stars is now possible,
and fully relativistic evolution codes are being used to study the
stability~\cite{sbs00,bss00} and gravitational collapse~\cite{s00,s03,ss02_1}
of such objects.
Binary neutron stars can now be evolved accurately for multiple
orbits~\cite{dmsb03,mdsb03}, and numerical simulations have been used to
study the final merger of these binaries~\cite{su00,stu03}.

The necessary tools are clearly being forged to enable numerical
relativity to model a wide variety of strong-field gravitational
phenomena.  Many interesting systems
in astrophysics involve the simultaneous presence of both black holes and
hydrodynamic matter fields, and these systems will require a code which can
handle both in order to model them reliably.

One important scenario involving both hydrodynamic matter and black holes
is core collapse in massive stars, an event
of immense importance due to its association with supernovae, the
formulation of neutron stars and black holes, and
gamma ray bursts (GRBs).  Some recent numerical simulations suggest~\cite{f99}
that a star must have mass less than about 20 $M_{\odot}$ for
core collapse to result in a conventional neutron star and
supernova explosion.  For progenitor masses between around $20 M_{\odot}$ and
$40 M_{\odot}$, the core collapses to a neutron star initially, but it eventually
implodes to a black hole, as ejected material slowly falls back onto
the remnant (see also~\cite{bzs00}).  For more massive stars, the core collapses
promptly to a black
hole. Such a massive system is a promising candidate for a GRB
\cite{mw99}.  There is growing
evidence that long duration GRBs are associated with
hypernovae that accompany the collapse of massive stellar cores. 
This evidence includes the association
of the low-energy GRB980425 with a supernova~\cite{g98}, the
presence of supernova-like features in the optical afterglow of
several GRBs~\cite{afterglow}, and the existence of freshly synthesized
elements in the ejecta of GRB 011211~\cite{r02}.  Most recently,
a hypernova was found to be temporally and spatially coincident
with a normal cosmological burst source, GRB 030329~\cite{h03}.  Most
models of the central engine of GRBs involve a black hole surrounded
by a rapidly accreting disk and a jet~\cite{GRB_reports}.  Three dimensional
fully relativistic simulations of both the black hole and the exterior matter
will be needed to test the feasibility
of various models for the production of GRBs from such ``collapsars'',
and it is likely that excision will be required to track the full evolution.

The merger of binary neutron stars is a promising source of
gravitational waves, as well as a prime candidate for
short duration GRBs~\cite{npp92}. 
Binary mergers of stars of high compaction collapse promptly to a
black hole~\cite{su00}.  The coalescence of
low-compaction neutron stars probably leads to the formation of
a hypermassive neutron star remnant~\cite{rs92,bss00,stu03} followed
by a delayed collapse~\cite{css03,lsd03}.  Either way,
excision of the black hole singularities is necessary to follow
binary mergers which form black holes with
accretion disks, emit gravitational waves, and drive
short-duration GRBs.

The dynamics of accretion flows onto a black hole is another problem of
great importance, since black holes are usually visible
electromagnetically only
through accretion.  When the mass of the accreting fluid is much
less than that of the black hole, then the matter can be evolved
on a fixed black hole spacetime.  When the masses
of the hole and the
matter are comparable, then a fixed black hole spacetime becomes a
bad approximation to the true metric, and the full
system must be evolved self-consistently.  This is particularly important
for determining if and when the disk may produce an instability, as in
the runaway instability~\cite{runaway}, or in the one-armed spiral
instability~\cite{one_arm} which can generate quasi-periodic gravitational
waves~\cite{sms1}.

Other coupled black hole-hydrodynamic matter systems include a neutron-star black hole
binary, which might be an important source of gravitational
radiation for LIGO, and also the disruption or capture of a
star by a massive black hole, which is expected to be a major source
of waves for LISA~\cite{bc03}.  Supermassive black hole seed formation
by the collapse of a massive or supermassive star is another important
example~\cite{SMBH,sms1}.

Successful attempts at evolving matter
and a black hole together in a dynamical spacetime using excision
have been rare.  Scheel {\it et al.}~\cite{sst95}
simulated the collapse of a spherically symmetric configuration of
collisionless matter in Brans-Dicke theory with excision.  The
spacetime within the
numerical domain was evolved until the appearance of
an apparent horizon.  At that time, an excision boundary was
introduced and the evolution of the exterior spacetime was continued.
An attempt to evolve a dynamical black hole spacetime with
hydrodynamic matter was undertaken by Brandt {\it et al}~\cite{bfims00}
in axisymmetry.  In that paper, a black hole is evolved with an
accretion flow, using the ADM formalism and
an isometry inner boundary condition at the apparent horizon. 
They were able to evolve several systems for up to
about 100$M$.  Little progress has been made since
then, presumably because the computational tools for performing
excision and relativistic hydrodynamics in 3+1 had to be perfected
independently.  We have only now reached the stage where these tools
can be put together successfully.

In this paper, we perform the first simulations which utilize
excision to evolve relativistic hydrodynamic matter in 3+1
dynamical spacetimes containing black holes.  In particular,
we present evolutions of the gravitational
collapse of stars from the beginning of collapse, through
black hole formation, to quiescent final states. 
We perform these
evolutions in two stages.  From the beginning of collapse
until the appearance of an apparent horizon, we evolve
using our new, relativistic hydrodynamics (BSSN) code without
excision (i.e. our ``pre-excision'' code).  After an apparent horizon
appears, we continue the evolution with a region inside the horizon
excised.  At the moment we introduce excision, a significant
amount of matter is still outside the excision zone and the
black hole is significantly distorted and in a nonstationary
state.  We follow its evolution to a final stationary state. 
In Section~\ref{method} we describe our evolution scheme,
including our gauge and boundary conditions.  In Section
\ref{diagnostics} we describe our code diagnostics. 
In Section~\ref{code_tests} we test our code on systems
with known behavior, including vacuum black holes, relativistic Bondi
accretion, Oppenheimer-Snyder collapse, and the collapse of
unstable polytropes, both non-rotating and rotating.  We
find that we are able to evolve many systems stably for hundreds of $M$. 
When we evolve systems with appreciable angular momentum, we
can only conserve $J$ for somewhat shorter durations, but this duration
can be extended by increasing resolution.  In Section
\ref{application} we apply our code to study the
late-time outcome of pressure-depletion-induced gravitational
collapse of rapidly-rotating polytropes with polytropic index
$n = 1$.  We find that stars
with $J/M^2<1$ collapse to Kerr black holes with no surrounding
disks.  Stars with $J/M^2>1$ collapse to tori, which then fragment. 
This fragmentation process can produce copious amounts of gravitational
radiation, originally referred to as ``splash radiation''~\cite{rrw74}. 
Finally, we summarize our results and discuss future improvements
to our code in Section~\ref{discussion}.

Throughout this paper, Latin and Greek indices denote spatial components
(1-3) and spacetime components (0-3), respectively.  We use geometrized
units, so that $G = c = 1$.

\section{Summary of Method}
\label{method}

Our basic code has been described in detail in previous
papers~\cite{dmsb03,ybs02}
and will be discussed here only briefly to
point out recent improvements.  Our code evolves the full Einstein
field equations coupled to relativistic hydrodynamics in 3+1
dimensions.  We have recently generalized this code using the
Cartoon methods of~\cite{abbhstt99,s00} so that it can perform
2+1 simulations in axisymmetry in the same coordinate system. 
In order to improve its behavior
near the intersection of the excision zone and the symmetry axis,
we add a small amount of Kreiss-Oliger
dissipation~\cite{chlp03,acst94,ko73} to
the evolution equation for the extrinsic curvature $\tilde A_{ij}$. 
A further description of
our axisymmetry algorithms, together with axisymmetry code tests, will
be presented in a forthcoming paper~\cite{lsd03}, in which the
effects of viscosity on differentially rotating binary neutron star
remnants are studied.

We evolve the field evolution equations using the BSSN formulation
\cite{sn95,bs99}.  In the BSSN system, one decomposes the 3-metric as
$\gamma_{ij} = e^{4\phi}\tilde\gamma_{ij}$ and the extrinsic
curvature as $K_{ij} = e^{4\phi}(\tilde A_{ij} + \tilde\gamma_{ij}K/3)$,
and one promotes the conformal connection coefficients
$\tilde\Gamma^i = -\tilde\gamma^{ij}{}_{,j}$ to independent
variables.  One then uses the ADM equations to write evolution
equations for the new set of fundamental variables: $\tilde\gamma_{ij}$,
$\phi$, $\tilde A_{ij}$, $K$, and $\tilde\Gamma^i$.  On each
time slice, these variables must satisfy the following constraint
equations:
\begin{eqnarray}
\label{Hamiltonian_BSSN}
  0 = \mathcal{H} &\equiv& 
                \tilde\gamma^{ij}\tilde D_i\tilde D_j e^{\phi}
                - {e^{\phi} \over 8}\tilde R  \\
		& & + {e^{5\phi}\over 8}\tilde A_{ij}\tilde A^{ij}
		    - {e^{5\phi}\over 12}K^2 + 2\pi e^{5\phi}\rho, 
	\nonumber \\
\label{momentum_BSSN}
  0 = {\mathcal{M}}^i &\equiv&
  \tilde D_j(e^{6\phi}\tilde A^{ji})- {2\over 3}e^{6\phi}\tilde D^i K
  - 8\pi e^{6\phi}S^i \\
\label{gamma_BSSN}
  0 = \mathcal{G}^i &\equiv& \tilde \Gamma^i + \tilde \gamma^{ij}_{~~,j} \\
\label{det_BSSN}
  0 = \mathcal{D} &\equiv& \det(\tilde\gamma_{ij}) - 1 \\
\label{trace_BSSN}
  0 = \mathcal{T} &\equiv& {\rm tr}(\tilde A_{ij})\ .
\end{eqnarray}
These constraints are solved only at the initial time, and
are used henceforward as diagnostics.  In an attempt to
improve the stability and accuracy of the BSSN formulation, one
can add multiples of the
above constraints to the field equations.  Many possible
modifications of this kind have recently been suggested
\cite{kst01,d87,ys01,kllpsst01,ys02,ybs02}.  We found a slight improvement
in ADM mass conservation by adopting the following modifications:
\begin{eqnarray}
\label{phimod}
\partial_t\phi &=& \cdots + c_{H1}\Delta T \alpha\mathcal{H} \\
\label{gmod}
\partial_t\tilde\gamma_{ij} &=& \cdots
  + c_{H2}\Delta T \alpha\tilde\gamma_{ij} \mathcal{H} \\
\label{Amod}
\partial_t\tilde A_{ij}  &=& \cdots
  - c_{H3}\Delta T \alpha\tilde A_{ij} \mathcal{H}\ ,
\end{eqnarray}
where $\Delta T$ is the timestep, $c_{H1} = 0.1$, $c_{H2} = 0.5$,
and $c_{H3} = 1$.  [For the complete right hand sides of
Eqs.~(\ref{phimod})-(\ref{Amod}), see~\cite{dmsb03}, equations
(12), (11), and (14).]  Modifications
similar to those in Eqs. (\ref{phimod}) and (\ref{gmod}) were suggested
in~\cite{ys02}, while a modification similar to Eq. (\ref{Amod}) has
recently been used in~\cite{am03} for doing excision in the ADM formulation
for pure vacuum spacetimes. 
Eq. (\ref{phimod}) introduces a diffusive term into the evolution of
$\phi$.  Eq. (\ref{Amod}) introduces a nonlinear damping term into the
evolution of $\tilde A_{ij}$.  We find that modification (\ref{Amod})
has the largest impact on accuracy.

Of crucial importance for
the stability of our code are our constraint additions to the
$\tilde\Gamma^i$ evolution equation.  As shown in~\cite{dmsb03}, our
equation for $\partial_t\tilde\Gamma^i$ has the terms
\begin{equation}
\label{evolve_Gamma}
\partial_t\tilde\Gamma^i =
  {2\over 3}\tilde\Gamma^i\beta^j{}_{,j}
  - \tilde\Gamma^j\beta^i{}_{,j} + \cdots
\end{equation}
Looking, for example, at the $x$-component of this equation,
\begin{equation}
\label{evolve_Gammax}
\partial_t\tilde\Gamma^x = 
   {2\over 3}\tilde\Gamma^x\beta^j{}_{,j}
  - \tilde\Gamma^x\beta^x{}_{,x} - \cdots
\end{equation}
we see that if $\beta^j{}_{,j}>0$ or $\beta^x{}_{,x}<0$, then
$\partial_t\tilde\Gamma^x$ contains a term
tending to produce exponential growth.  We lessen the possibility of an
instability caused by these terms by using (\ref{gamma_BSSN}) to replace
(\ref{evolve_Gammax}) with
\begin{eqnarray}
\nonumber
\partial_t\tilde\Gamma^x &=& 
  {2\over 3}\left[\beta^j{}_{,j} + \lambda_A\left|\beta^j{}_{,j}\right|\right]
  (-\tilde\gamma^{xk}{}_{,k}) - {2\over 3}\lambda_A\left|\beta^j{}_{,j}\right|\tilde\Gamma^x \\
\nonumber
  & &-\left[\beta^x{}_{,x} + \lambda_B\left|\beta^x{}_{,x}\right|\right]
  (-\tilde\gamma^{xk}{}_{,k}) - \lambda_B\left|\beta^x{}_{,x}\right|\tilde\Gamma^x \\
\label{modify_Gamma}
  & & + \cdots
  \ ,
\end{eqnarray}
and similarly for $\tilde\Gamma^y$ and $\tilde\Gamma^z$.  Note
that the ``exponential'' terms in the above equation (i.e. the
terms proportional to $\tilde\Gamma^x$) are now guaranteed to
be exponential {\it decay} terms.  We find
good results with $\lambda_A = 2/3$ and $\lambda_B = 3/4$.

Alcubierre {\it et al.}~\cite{aabss00} find improved behavior when they
enforce the constraint ${\mathcal{T}} = 0$.  Yo {\it et al.}~\cite{ybs02}
found it useful to enforce ${\mathcal{T}} = {\mathcal{D}} = 0$. 
We instead apply the reasoning above to modify the evolution
equations for $\tilde\gamma_{ij}$ and $\tilde A_{ij}$.  Thus,
in the equation for $\tilde\gamma_{xx}$, we find the terms
\begin{equation}
\partial_t\tilde\gamma_{xx} = \left(-{2\over 3}\beta^j{}_{,j}
  + 2\beta^x{}_{,x}\right)\tilde\gamma_{xx} + \cdots
\end{equation}
which we replace by
\begin{eqnarray}
\nonumber
\partial_t\tilde\gamma_{xx} &=& 
  {2\over 3}\left[-\beta^j{}_{,j} + \lambda_C\left|\beta^j{}_{,j}\right|\right] G_{xx}
  - {2\over 3} \lambda_C \left|\beta^j{}_{,j}\right|\tilde\gamma_{xx} \\
\nonumber
  & &+ 2\left[\beta^x{}_{,x} + \lambda_D\left|\beta^x{}_{,x}\right|\right]G_{xx}
  - 2 \lambda_D \left|\beta^x{}_{,x}\right|\tilde\gamma_{xx} \\
  & & + \cdots \ ,
\end{eqnarray}
where $G_{xx}$ is the value of $\tilde\gamma_{xx}$ as computed from
the five other independent components of $\tilde\gamma_{ij}$, assuming
${\mathcal{D}} = 0$.  We perform the same substitution for
$\tilde\gamma_{yy}$ and $\tilde\gamma_{zz}$.  We use $\lambda_C = 2/3$
and $\lambda_D = 1/10$.  In a similar fashion, we modify the evolution
of $\tilde A_{xx}$, $\tilde A_{yy}$, and $\tilde A_{zz}$
from
\begin{equation}
\partial_t\tilde A_{xx} = \cdots + \left(-{2\over 3}\beta^j{}_{,j}
  + 2\beta^x{}_{,x} + \alpha K\right)\tilde A_{xx}
\end{equation}
to
\begin{eqnarray}
\nonumber
\partial_t\tilde A_{xx} &=&
  {2\over 3}\left[-\beta^j{}_{,j} + \lambda_C\left|\beta^j{}_{,j}\right|\right] H_{xx}
  - {2\over 3} \lambda_C \left|\beta^j{}_{,j}\right|\tilde A_{xx} \\
\nonumber
  & &+ 2\left[\beta^x{}_{,x} + \lambda_D\left|\beta^x{}_{,x}\right|\right]H_{xx}
  - 2 \lambda_D \left|\beta^x{}_{,x}\right|\tilde A_{xx} \\
\nonumber
  & &+ \left[\alpha K + \lambda_E\left|\alpha K\right|\right]H_{xx}
  - 2 \lambda_E \left|\alpha K\right|\tilde A_{xx} \\
  & &+ \cdots\ ,
\end{eqnarray}
and similarly for the other two components. 
Here $\lambda_E = 0.1$, $\lambda_C$ and $\lambda_D$ are the same as
above, and $H_{xx}$ is the value of $\tilde A_{xx}$
computed from the five other independent components of $\tilde A_{ij}$
assuming ${\mathcal{T}} = 0$.

We take spatial derivatives in a centered way---we do not use
causal differencing.  The only exception, as suggested by~\cite{ab01},
is in the advection terms along the shift $\beta^i\partial_i$, for
which we use the second-order upwind differencing described in
\cite{s99_2}.

Our hydrodynamics scheme uses van-Leer type
advection and artificial viscosity shock handling~\cite{dmsb03}. 
It is known that such schemes can be inaccurate for ultrarelativistic
flows~\cite{f03}.  We monitor the Lorentz factors
of our fluids, and find that they never exceed $\approx$2, which
is around the upper limit for accurate evolutions with a van Leer
code.  In addition, most of our runs do not involve strong shocks. 
We thus believe that our hydrodynamics scheme is adequate for
the present purposes, although we eventually may have to
improve it.  Our hydrodynamics scheme employs the ``no atmosphere''
approach~\cite{dmsb03}, so that the density at any point on our grid
is allowed to fall to zero.  It is important that we
are able to dispense with an artificial atmosphere.  If we could
not, then in situations where all the matter in the problem falls
into the black hole, the hole would continue to accrete atmosphere
indefinitely, and its mass would continue to grow unphysically.

The boundary conditions we apply at the edge of the excision zone are
described in detail in~\cite{ybs02}.  They consist of taking the time
derivatives of quantities at the excision boundary from the time
derivatives of these quantities at adjacent points.  We use spherical
excision regions inside the apparent horizon throughout
(see~\cite{cn03} regarding the superiority of spherical to cubic excision
regions).  We have
tried several boundary conditions for the matter variables, and
have found that our results are insensitive to the choice, as they
should be.  In the runs described below, we simply set the matter
variables equal to zero when they hit the excision zone, thus
making the excision boundary a perfect one-way membrane.

The lapse and shift must be chosen in such a way that the total
system of evolution equations is stable.  It is also desirable that
the gauge conditions are chosen so that, as the system settles into
equilibrium, it appears stationary in the adopted coordinates. 
We have experimented with several choices for the lapse $\alpha$
and shift $\beta^i$, and we have found that driver conditions
using the second time derivatives of $\alpha$ and
$\beta^i$ provide the most stable evolutions. 
Following the suggestion of Alcubbierre {\it et al}~\cite{abpst01}, we
have had great success with the hyperbolic shift driver condition:
\begin{equation}
\label{hb_shift}
\partial^2_t\beta^i = b_1(\alpha\partial_t\tilde\Gamma^i
 - b_2\partial_t\beta^i)\ ,
\end{equation}
with $b_1 = 0.75$ and $b_2 = 0.27 M^{-1}$ (c.f.~\cite{ab01}). 
One can create a hyperbolic lapse condition by introducing
two coupled first-order equations and a new function ${\cal A}$
\begin{eqnarray}
\label{hb_lapse_nok3}
\nonumber
\partial_t \alpha &=& \alpha {\cal A} \\
\partial_t {\cal A} &=& -a_1(\alpha \partial_tK + a_2\partial_t\alpha)\ ,
\end{eqnarray}
with $a_1 = 0.75$ and $a_2 = 0.27 M^{-1}$. 
The $\alpha$ in front of ${\cal A}$ in the first equation is a ``safety''
feature, to prevent the lapse from dropping to zero.  With
this safety feature, we find that the lapse levels off at finite
positive values everywhere on and outside the excision zone for
all our runs, thereby maintaining a ``horizon penetrating''
($\alpha > 0$) time coordinate.  However, at late times ($t\sim 200 M$),
we find that the asymptotic values of some of our variables
(e.g. $\tilde\gamma_{xx}$) begin to drift, increasing linearly
with time. 
This drift is also present when harmonic slicing, another
slicing with a hyperbolic character~\cite{a03}, is adopted.  Apparently,
Eq.(\ref{hb_lapse_nok3}) does not sufficiently restrict the coordinate
system's evolution.  We remove the drift by
adding a third term to Eq.(\ref{hb_lapse_nok3}) proportional to
$K - K_{\rm drive}$, where $K_{\rm drive}$ is some reasonable
positive function.  In this way, the value of $K$ itself, and
not just its time derivative, is ``driven''.  We shall
refer to this slicing as our ``hyperbolic lapse''.  The complete
slicing condition is
\begin{eqnarray}
\label{hb_lapse}
\nonumber
\partial_t \alpha &=& \alpha {\cal A} \\
\partial_t {\cal A} &=& -a_1(\alpha\partial_tK \\
\nonumber
   & &\ + a_2[\partial_t\alpha + e^{-4\phi}\alpha(K-K_{\rm drive})])\ .
\end{eqnarray}
Here the $e^{-4\phi}\alpha$ factor is chosen so that the new term is
small in the strong-field region, where (\ref{hb_lapse_nok3}) works
well, but becomes comparable to the other terms in the outer portions
of the grid, where it successfully removes the drift.

We have tried several forms for $K_{\rm drive}$.  The simplest, and
usually adequate choice, is zero.  This drives $K$ to zero (maximal
slicing) and usually causes a very slow downward drift in the lapse
near the horizon.  For many astrophysical applications, where we only need to
evolve for several hundred $M$, this is usually unimportant.  However, the
effect can be removed by a better choice of $K_{\rm drive}$. 
One possibility is $K_{\rm init}$, the value of $K$ at the time excision
is introduced.  Another choice is $K_{KS}$, a function whose form is
inspired by the Kerr-Schild representation of a Kerr black hole
[c.f. Eq.~(36) of~\cite{ybs02}].
\begin{eqnarray}
\label{K_KS}
K_{KS}(\alpha,\beta^i) &=& 2\alpha^3(1+H)l^i H_{,i} + 2\alpha H l^i{}_{,i} \\
\nonumber
H &=& {1\over 2}(\alpha^{-2} - 1) \\
\nonumber
l^i &=& \beta^i / (2\alpha^2H)
\end{eqnarray}
Note that when we choose this functional form for $K_{KS}$, the lapse and
shift typically are not the same as the Kerr-Schild $\alpha$ and $\beta^i$.

For $K = K_{\rm drive}$, we apply our usual excision boundary
conditions on $\alpha$.  Otherwise, there are no spatial derivatives
in Eq. (\ref{hb_lapse}), and no explicit inner boundary condition is
needed.  In some cases, however, we have found more accurate results
when we hold the values of the lapse on the excision zone fixed in time
(the ``frozen'' inner boundary condition).

\section{Diagnostics}
\label{diagnostics}
Our most important diagnostics are the conserved
mass $M$ and angular momentum $J$.  These are both defined by
surface integrals at infinity~\cite{MandJ}:
\begin{eqnarray}
\label{Mdef}
M &=& {1\over 16\pi}\int_{r=\infty}\sqrt{\gamma}\gamma^{im}\gamma^{jn}
                                 (\gamma_{mn,j} - \gamma_{jn,m}) d^2S_i \\
\label{Jdef}
 J_i &=& {{1}\over{8\pi}}\varepsilon_{ij}{}^k \int_{r=\infty}
                         x^j K_k^m d^2S_m.
\end{eqnarray}
We measure $M$ and $J$ by applying Gauss' Law to obtain a surface
integral over an inner surface, $\partial\Omega$, (which encloses the
singularity) plus
a volume integral over the space outside this
surface, $\Omega$.  Details of this calculation are presented
in~\cite{ybs02}.  The final integrals are
\begin{eqnarray}
   M&=& \frac{1}{16\pi}\int_\Omega d^3x\bigg[e^{5\phi}
        \bigg(16\pi\rho + \tilde{A}_{ij}\tilde{A}^{ij} 
	-\frac{2}{3}K^2\bigg)\nonumber\\
    &&\qquad\qquad\qquad\qquad -\tilde{\Gamma}^{ijk}\tilde{\Gamma}_{jik}
    + (1-e^\phi)\tilde{R}\bigg]\nonumber\\ 
    &&+ \frac{1}{16\pi} \oint_{\partial\Omega} 
    (\tilde{\Gamma}^i-8\tilde{D}^ie^\phi) d\tilde{S}_i\label{Mivp} \\
   J_i&=&\frac{1}{8\pi} \epsilon_{ij}{}^k\int_\Omega
           \bigg[e^{6\phi}\bigg(\tilde{A}^j{}_k 
           + \frac{2}{3}x^j\tilde{D}_kK\nonumber\\
    &&\qquad\qquad\quad- \frac{1}{2} x^j\tilde{A}_{ln}
     \partial_k\tilde{\gamma}^{ln}{}+8\pi x^js_k\bigg)\bigg]d^3x\nonumber\\
    &&+\frac{1}{8\pi} \epsilon_{ij}{}^k \oint_{\partial\Omega}
           e^{6\phi} x^j \tilde{A}^l{}_k d\tilde{S}_l.\label{Aivp}\ .
\end{eqnarray}
We choose the inner surface $\partial\Omega$ to be a sphere with a
coordinate radius about twice that of the excision boundary.  This
puts $\partial\Omega$ slightly outside the apparent horizon in the
simulations reported below.

In our pre-excision code, $\Omega$ is chosen to cover the entire
numerical grid, and there is no surface integral contribution.  The
rest mass $M_0$ cannot be used as a diagnostic because it is
conserved identically in our pre-excision code.  Our pre-excision
code also conserves $J$ identically in axisymmetry~\cite{lsd03}.  With excision,
$M_0$ is not expected to be conserved in $\Omega$, since matter falls into the
excision region.  When evolving with excision, $J$ is not identically
conserved, even in axisymmetry, and thus serves as a code check together
with $M$.

Once a black hole is present, we detect it by using an apparent horizon
finder (see~\cite{bcsst96} for details).  As the system approaches
stationarity, the apparent horizon will approach the event horizon. 
We estimate the size the horizon in our coordinate system by the
radius $r_{\rm AH}$ constructed from the $l = 0$, $m = 0$ moment
of the horizon surface. 
From the surface area of the apparent horizon, we compute the
irreducible mass $M_{\rm irr}$ defined by
\begin{equation}
M_{\rm irr} = \sqrt{{\mathcal{A}}/16\pi^2}
\end{equation}
We also compute the proper circumference of the
horizon in the equatorial $(xy)$ plane, which we call $C_{\rm eq}$,
and we compute the proper circumference
in the meridional $(xz)$ plane, which we call $C_{\rm pol}$. 
For static nonrotating black holes,
$C_{\rm eq} = C_{\rm pol} = 4\pi M$.  For stationary rotating black
holes, one can compute $C_{\rm eq}$ and $C_{\rm pol}$ from the
Kerr metric in Boyer-Lindquist coordinates to be
\begin{eqnarray}
  C_{\rm eq}  &=& 4\pi M \label{C_eq} \\
  C_{\rm pol} &=& 4 M \int_0^{\pi/2} d\theta
                  \sqrt{2 + 2\sqrt{1-q^2} + q^2\sin^2\theta}\ , \label{C_pol}
\end{eqnarray}
where $q \equiv J/M^2$ is the spin parameter of the black hole. 
The ratio $C_{\rm pol}/C_{\rm eq}$ varies from
1 for $q = 0$ to 0.6 for $q = 1$.  For the black holes in our
simulations, we infer
the horizon mass $M_{\rm AH}$ from $C_{\rm eq}$ and Eq.~(\ref{C_eq}). 
We infer the horizon angular momentum $J_{\rm AH}$ from
$C_{\rm pol}/C_{\rm eq}$ and Eq.~(\ref{C_pol}), together with $M_{\rm AH}$.

Finally, we find the ergosurface of the black hole.  The ergosphere
is defined in the stationary limit, in which case
$\partial\over\partial t$ is a Killing vector, and the ergosurface
is defined as the surface where
$g_{00} = {\partial\over\partial t}\cdot{\partial\over\partial t} = 0$,
with $g_{00}>0$ inside and $g_{00}<0$ outside.

As in~\cite{ab01,ybs02}, we gauge the degree to which a field $f$ reaches
stationarity by monitoring $\Delta f(t)$, defined to be the L2 norm
of $f(t)-f(t-\Delta T)$, where $\Delta T$ is the timestep.  We compute
the L2 norm of a gridfuntion $g$ by summing over every gridpoint $i$:
\begin{equation}
  L2(g) = \sqrt{\sum_i g_i^2}
  \label{L2}
\end{equation}

\section{Tests}
\label{code_tests}

\subsection{Field Code Test:  Vacuum Black Holes}

\begin{figure}
\epsfxsize=2.8in
\begin{center}
\leavevmode \epsffile{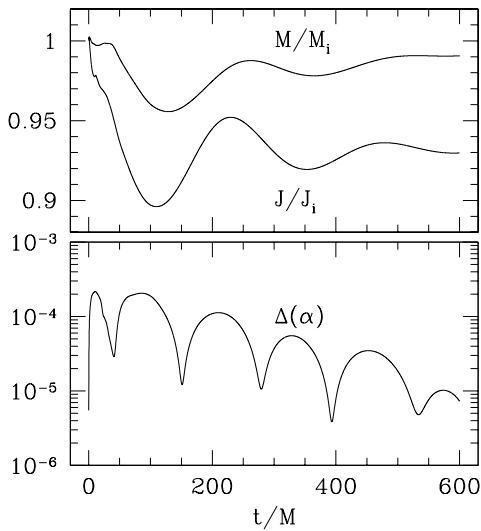}
\end{center}
\caption{ The evolution of the mass $M$, angular momentum $J$, and
  lapse variation $\Delta\alpha$ for the evolution of an $a/M = 0.4$
  black hole in Kerr-Schild coordinates.  We use a $30^2$ grid to cover
  the meridional plane.. }
\label{ks_fig}
\end{figure}

In a previous paper~\cite{ybs02}, we used our code to evolve
isolated, stationary black hole spacetimes in Kerr-Schild
coordinates.  These coordinates have the advantages of being
horizon-penetrating ($\alpha\neq 0$ at the horizon) and
providing a manifestly stationary metric.  We were able to
evolve both stationary and rotating black holes for
arbitrarily long times.  We succeeded in doing this both when
evolving only one octant of the space and when evolving
the full space without any symmetry assumptions.  These evolutions
were done in three dimensions using a different set of gauge
conditions from those utilized in this paper.  In Figure
\ref{ks_fig}, we show the evolution of a $a/M = J/M^2 = 0.4$ Kerr black
hole in Kerr-Schild coordinates using our 2D axisymmetry code
and our hyperbolic gauge conditions.  For this case, we use
a frozen inner boundary condition on $\alpha$, and turn off
the third term in (\ref{hb_lapse}).  (Using
$K_{\rm drive} = K_{\rm init}$ gives similar results.)  We use
a grid spacing of $\Delta X = 0.4 M$, with outer boundaries at
$12 M$ and an excision zone at a coordinate radius of $1.5 M$,
as was used by~\cite{ybs02}.  The event horizon for
$a/M = 0.4$ is located at $r_{\rm eq} = 1.917 M$ in these coordinates.

As a second test, adapted from~\cite{ab01}, we evolve a Schwarzschild
black hole in initially isotropic coordinates.  Choosing $\alpha = 1$
and $\beta^i = 0$ at $t = 0$, the initial metric is
\begin{equation}
ds^2 = -dt^2 + \left(1+{2M\over r}\right)(dx^2 + dy^2 + dz^2)\ ,
\end{equation}
where $r = \sqrt{x^2 + y^2 + z^2}$.  The event horizon is
located at $r = 0.5 M$ in these coordinates.  Physically, this black hole
is stationary, but it does not appear stationary in the coordinates
generated by Eqns.~(\ref{hb_shift}) and~(\ref{hb_lapse})
starting with the initial
lapse and shift cited above.  By evolving this spacetime, we
check that our excision code can work with coordinates other
than stationary Kerr-Schild.  We also check the ability of
our gauge conditions to ``find'' coordinate systems which make
the metric manifestly stationary.  We allow the
lapse to drop, so we do not freeze the lapse at the excision zone,
but employ equation (\ref{hb_lapse}) everywhere.  We use
$K_{\rm drive} = K_{\rm init} = 0$, since $K_{KS}$ is singular
for our value of $\alpha$ at $t = 0$ [see equation (\ref{K_KS})].

\begin{figure}
\epsfxsize=2.8in
\begin{center}
\leavevmode \epsffile{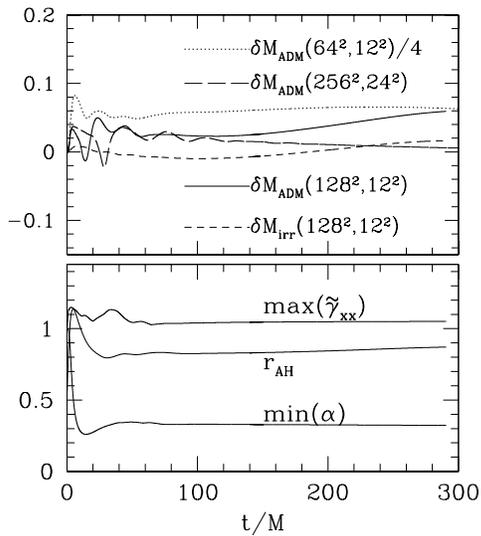}
\end{center}
\caption{ The evolution of a nonrotating black hole in our hyperbolic
  gauges, starting in isotropic coordinates with $\alpha = 1$,
  $\beta^i = 0$.  On top, we show the deviations of the ADM mass
  $M_{\rm ADM}$ and the irreducible mass $M_{\rm irr}$ from their
  initial value:  $\delta M = (M - M_i)/M_i$.  $\delta M$ is shown
  for runs with outer boundaries at $12 M_i$ using a $128^2$ grid
  and using a $64^2$ grid, to demonstrate convergence.  We also show
  a run with outer boundaries at $24 M_i$ using a $256^2$ grid to
  determine the effect of the outer boundary.  Below, we show
  the time evolution on the $128^2$ grid of the apparent horizon
  coordinate radius $r_{\rm AH}$ and the maximum values of $\alpha$
  and $\tilde\gamma_{xx}$ on the grid. }
\label{ss_fig}
\end{figure}

In Figure~\ref{ss_fig}, we plot the results for a run in
axisymmetry with outer boundaries at $12 M$, an excision
radius of $0.36 M$, and a grid of $128^2$ to cover the
meridional ($xz$) plane.  Also shown are scaled results for
a $64^2$ run to demonstrate convergence.  We also performed
a run on a $256^2$ grid with the same resolution as the $128^2$
run but with the
outer boundaries at $24 M$.  From the figure, we see that the
error can be controlled by the grid resolution and the location
of the outer boundaries. 
We see that the surface area of the apparent horizon (i.e. $M_{\rm irr}$)
remains nearly constant while the coordinates adjust to create a stationary
system.  This indicates that the apparent horizon is following the
event horizon well.  The coordinate adjustment is reflected in the
initial increase in the coordinate radius of the horizon and in the
drop of the lapse.  Note that the lapse settles quickly, and that it
remains positive everywhere outside and at the excision zone.  To
check that the black hole remains a Schwarzschild black hole, we
monitor $C_{\rm eq}$ and $C_{\rm pol}$ and find that they both
remain equal to $4 \pi M$ to within one percent.

\subsection{Hydro Code Test:  Relativistic Bondi Flow}
Next, we test our hydrodynamics code by solving an accretion problem
that has an exact solution. 
In a previous paper~\cite{dmsb03}, we confirmed our code's ability
to accurately simulate shocks, spherical dust collapse,
nonrotating and rotating polytropes, and binary polytropes. 
Now we test its ability to maintain stationary, adiabatic, spherically
symmetric accretion onto a Schwarzschild black hole, in accord
with the
relativistic Bondi accretion solution for $\Gamma = 1.5$~\cite{st83}. 
Following the suggestion of~\cite{pf98},
we write the metric in Kerr-Schild (ingoing Eddington-Finklestein)
coordinates; in this way, all the variables are
well behaved at the horizon.  We begin by holding the field
variables fixed in order to prevent the black hole from
growing due to accretion.

\begin{figure}
\epsfxsize=2.8in
\begin{center}
\leavevmode \epsffile{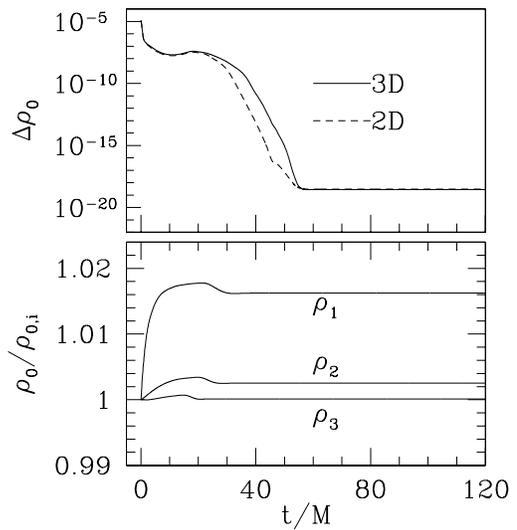}
\end{center}
\caption{ The settling of the rest-mass density to steady-state,
  starting from the analytic
  value.  The change per timestep quickly drops to the machine level. On
  top, we plot $\Delta\rho_0$ for both the $64^2$ 2D run and the $64^3$
  3D run.  Below, we show the time evolution of $\rho_0$ at three points
  on the diagonal line $x = y = z$ in the 3D run, each normalized to its
  initial value.  $\rho_1$ corresponds to
  $\rho_0$ measured at $r = \sqrt{3}x = 2 M$, $\rho_2$ to $\rho_0$
  at $r = 6 M$, and $\rho_3$ to $\rho_0$ at $r = 10 M$. }
\label{bondi_fig}
\end{figure}

We evolve this system twice, once using a $64^2$ grid in 2+1 and
once using a $64^3$ grid in 3+1.  We place
outer boundaries at $12 M$ and an excision zone at a
coordinate (areal) radius of $1.5 M$.  At $t = 0$, we set the
density and velocity profiles according to the exact solution
for $\Gamma = 1.5$ and an accretion rate $dM/dt|_{\rm acc} = 0.0031$,
with a sonic radius
at $10^5 M$. This accretion rate is maintained through the evolution
by fixing the hydrodynamic variables on the outer boundaries at their
exact steady-state values. 
In Figure~\ref{bondi_fig}, we plot $\Delta(\rho_0)$, as defined in
Section~\ref{diagnostics}, and also the values of $\rho_0$ at selected
points in the accretion flow.  For $\Delta(\rho_0)$, we reach
machine precision after less than $100 M$, making further
integration unnecessary~\cite{ybs02}.  (The velocity fields have also
frozen near their initial values by this time.)

When we allow the fields to evolve, we see the irreducible
mass of the hole grow at a rate
$dM_{\rm irr}/dt \approx 0.9 dM/dt|_{\rm acc}$.  This error is consistent
with the errors in our irreducible mass found  at this numerical
resolution, even in the absence of accreting matter.

\subsection{Oppenheimer-Snyder Collapse}

\begin{figure}
\epsfxsize=2.8in
\begin{center}
\leavevmode \epsffile{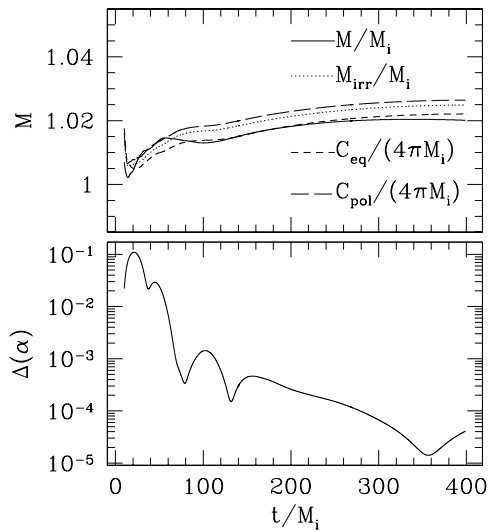}
\end{center}
\caption{ ADM mass, horizon diagnostics, and $\Delta\alpha$ for the
  collapse of a homogeneous sphere of dust to a Schwarzschild black
  hole.  Collapse begins at $t = 0$ and black hole excision occurs at
  $t = 10 M$. }
\label{os_fig}
\end{figure}

\begin{figure}
\epsfxsize=2.8in
\begin{center}
\leavevmode \epsffile{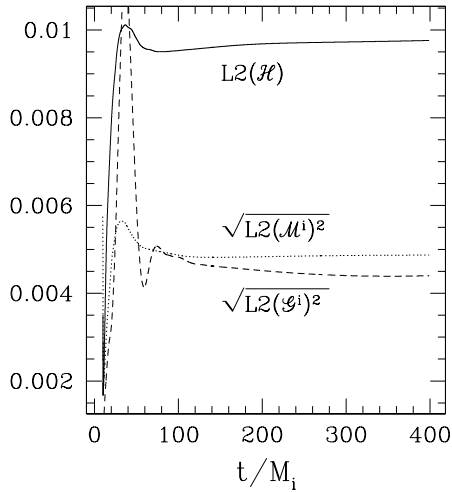}
\end{center}
\caption{ Violation of the Hamiltonian ${\mathcal H}$, momentum
  ${\mathcal M}^i$, and Gamma ${\mathcal G}^i$ constraints as a
  function of time for the collapse depicted in Fig.~\ref{os_fig}. 
  We plot the un-normalized L2 norms, where we use the shorthand
  $L2(\mathcal{M}^i)^2 =
  L2(\mathcal{M}^x)^2 + L2(\mathcal{M}^y)^2 + L2(\mathcal{M}^z)^2$ and \\
  $L2(\mathcal{G}^i)^2 =
  L2(\mathcal{G}^x)^2 + L2(\mathcal{G}^y)^2 + L2(\mathcal{G}^z)^2$.
}
\label{oscon_fig}
\end{figure}

Next, we simulate the Oppenheimer-Snyder collapse of a homogeneous
spherical ball of dust to a black hole.  The
behavior of this system is known in several coordinate systems
\cite{os39,pst85,pst86}.  We use a $160^2$ grid with outer boundaries at $14 M$. 
At $t=0$, the dust is at rest and has an areal radius
of $3 M$.  We start in an isotropic coordinate system, in which
$\tilde\gamma_{ij} = \delta_{ij}$.  Our initial $\alpha$ and $\beta^i$
are set by enforcing maximal slicing and the minimal distortion gauge
condition, respectively
%and set $\alpha$ and $\beta^i$
%from the conditions $K = \partial_tK = 0$ and
%$\tilde\Gamma^i = \partial_t\tilde\Gamma^i = 0$, respectively
(see~\cite{pst85}).
Since the ball has no pressure support, it immediately
begins to collapse.  During the first phase of this collapse, there
are no trapped regions and no singularities, so we evolve the entire
grid without excision.  Our code checks during this
part of the evolution are well satisfied; see~\cite{dmsb03,lsd03}. 
For gauge conditions during this
no-excision phase, we use our hyperbolic lapse and shift drivers. 
We evolve in this way from
$t = 0$ to $t = 11 M$, at which point our no-excision code crashes due to
its inability to resolve the central region (``grid stretching''). 
An apparent horizon
appears at $t = 9 M$ at a coordinate radius of $r_{\rm AH} = 0.96 M$
with an irreducible
mass of $M_{\rm irr} = 1.02 M$.  We next repeat the evolution from
$t = 10 M$ with our excision algorithm and
an excision boundary at radius $r_{\rm ex} = 0.7 M$.  At this
point, only 1.2\% of the rest mass is outside the horizon,
but the spacetime in our coordinates is still changing.  We
continue to evolve with our hyperbolic gauges, and we allow $\alpha$
to drop at the excision boundary.  In this example,
using $K_{\rm drive} = K_{\rm KS}$ is far superior to any other
choice, since only then does the lapse settle quickly.  As
we continue the evolution, the
remaining exterior rest mass falls into the excision zone over the course
of the next $100 M$, and we are left with a vacuum spacetime.  We
evolve for $400 M$, by which time the system has long since settled to
a Schwarzschild black hole.  Oppenheimer-Snyder collapse does have an
analytic solution in Friedmann coordinates,
but not in the coordinates we are using, which are defined by
our gauge conditions (\ref{hb_shift}) and (\ref{hb_lapse}) together
with the boundary conditions on $\alpha$ and $\beta^i$ at $r_{\rm ex}$. 
Therefore we check the accuracy of our evolution using global invariants. 
In Figure~\ref{os_fig}, we show our mass diagnostics for the
post-excision run, which confirm
that the end product is a Schwarzschild black hole, and we plot
$\Delta\alpha$ as proof of stationarity.  In  Figure~\ref{oscon_fig},
we plot the magnitude of the constraint violations as functions of
time.  These show that the error is not growing during the long
stationary evolution.

\subsection{Collapse of a TOV Star}
The previous example possessed spherical symmetry and no pressure. 
In our next test, we study the collapse of an unstable nonrotating,
spherical polytrope, whose initial state is given by the solution to
the TOV equations~\cite{ov39}.
\begin{figure}
\epsfxsize=2.8in
\begin{center}
\leavevmode \epsffile{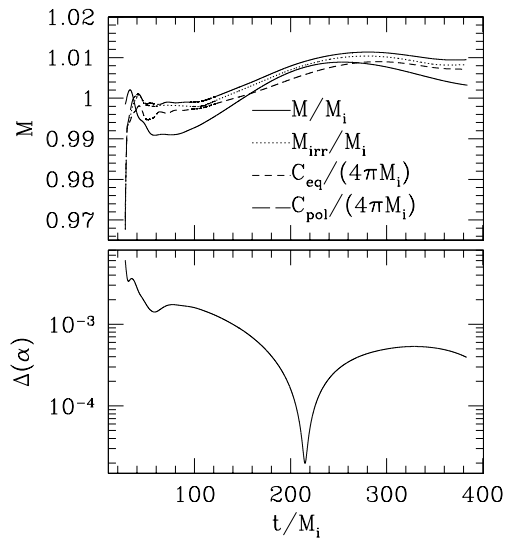}
\end{center}
\caption{ ADM mass, horizon diagnostics, and $\Delta\alpha$ for the
  collapse of a nonrotating, unstable $n = 1$ polytrope from apparent horizon
  formation at $t/M = 27$ through final stationarity.  The code is axisymmetric
  and uses a $128^2$ grid. }
\label{tov_fig}
\end{figure}

For initial data, we take a perfect fluid with equation of state
$P = \kappa \rho_0{}^{1+1/n}$, with $n = 1$, and we choose our
units such that $\kappa = 1$~\cite{foot1}.  In these units,
the $n = 1$ TOV sequence has a turning point at the critical
central rest density $\rho_c^{\rm crit} = 0.32$ where the
ADM mass of the star is $M_{\rm max} = 0.164$.  We choose to evolve
a star with initial central rest density $\rho_c = 0.5$ and ADM
mass $M = 0.158$.  As this star is on the unstable branch of the
$n = 1$ sequence,
it is unstable to radial oscillations and will collapse to a black
hole.  We evolve the first part of the collapse without excision
using a $128^2$ grid, with outer boundaries at $12.7 M$
and with our hyperbolic drivers.  We evolve from $t = 0$ to
$t = 28.5 M$, locating an apparent horizon at $t = 27 M$ with
radius $r = 0.6 M$ and irreducible mass $M_{\rm irr} = 0.95 M$.  We begin
an excision run from $t = 27.8 M$, at which point 4\% of the rest
mass is still outside the apparent horizon and 8\% is outside of
the excision zone.  All of this matter falls into the excision zone
by $t = 31.6 M$.  It should be emphasized that the spacetime in these
coordinates is more dynamical than the above numbers might suggest:
e.g. during the first $10 M$ of post-excision evolution, the maximum value of
$\tilde A^{ij}\tilde A_{ij} M^2$ increases from 0.25 to 0.44. 
%the maximum value of $K^2 M^2$ from 0.012 to 0.031. 
The system settles quickly
thereafter, as we see by evolving an additional $350 M$ to $390 M$. 
In Figure \ref{tov_fig}, we show our diagnostics for this run.
\begin{figure}
\epsfxsize=2.8in
\begin{center}
\leavevmode \epsffile{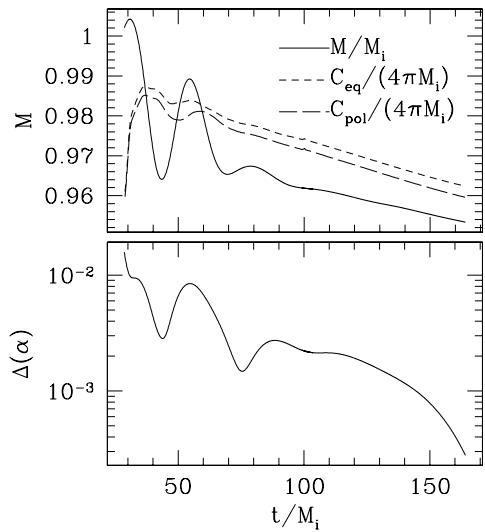}
\end{center}
\caption{ Same as for Figure~\ref{tov_fig}, but now the collapse is simulated
  on a 3D $64^3$ grid.}
\label{3Dtov_fig}
\end{figure}

All the runs described above were carried out on two-dimensional
axisymmetric grids. 
In Figure~\ref{3Dtov_fig}, we show diagnostics for the same collapse
in a three dimensional simulation, with a $64^3$ grid and boundaries
at $[0,12.7 M]^3$ (employing octant symmetry to evolve only the upper octant). 
The behavior of each quantity is similar to that in the 2D run.

\newcommand{\stara}{A}
\newcommand{\starb}{B}
\newcommand{\starc}{C}

\subsection{Collapse of a Rotating Star}
\label{rot_collapse}

Gravitational collapse of astrophysically realistic stars will
involve rotation. 
Even if the progenitor star rotates slowly, it will spin up as
it collapses if it conserves angular momentum.  It is therefore
important to test our code by simulating the collapse of a rapidly
rotating star.
\begin{figure}
\epsfxsize=2.8in
\begin{center}
\leavevmode \epsffile{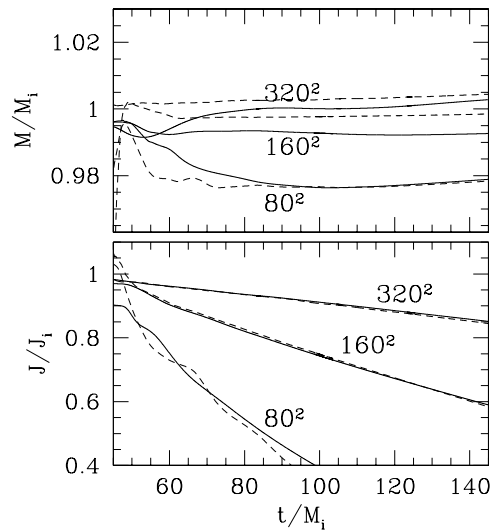}
\end{center}
\caption{ Mass $M$ and angular momentum $J$ during the post-excision
  phase of the collapse
  of star \stara.  We show results for axisymmetric runs carried
  out with a $80^2$, a $160^2$, and a $320^2$ grid.  Both $M$
  and $J$ are measured in two ways.  The solid lines are quantities
  as measured by the integrals (\ref{Mivp}) and (\ref{Aivp}).  The
  dashed lines are obtained by measuring the geometry of the apparent
  horizon and comparing with the Kerr metric  ($M_{\rm AH}$ as inferred by
  $C_{\rm eq}$ and $J_{\rm AH}$ by $C_{\rm pol}/C_{\rm eq}$).  For
  the $160^2$ and $320^2$ runs, the two $J$ measurements lie on top of
  one another. }
\label{rot_fig}
\end{figure}

The star we adopt as initial data, labeled \stara, is described in
Table~\ref{star_table}.  The initial data was
obtained using the relativistic equilibrium code of~\cite{cst92}. 
Star \stara\ is a ``hypermassive star'' with
a mass $M = 0.19$, which is 20\% higher than $M_{\rm max}$, the
maximum allowed mass of a nonrotating TOV star.  Star \stara\ is able
to maintain
this mass because of the added support against gravity provided by
(differential) rotation.  The star has $J/M^2 = 0.57$, so that the eventual Kerr
hole will have appreciable spin, assuming all of the mass and angular
momentum is captured by the hole.  Even prior to collapse, the
effects of angular momentum on the star are significant, as we can
see by noting that the radius of the star on the rotation $z$-axis (the
polar radius) is only 70\% of the radius of the star in the
equatorial plane (the equatorial radius).  Star \stara\ has a
differential rotation profile (see next section), so there are no
turning-point theorems which can be applied to determine the stability
of this star, but we find numerically that it is unstable to
collapse.  Perturbations due to numerical (roundoff) error are sufficient
to trigger the collapse, but
the onset timescale for collapse is not independent of resolution. 
In order to do convergence studies, we deplete a small percentage
(4\%) of the initial pressure, so that the initial perturbation is
resolution-independent.  This perturbation is so small that
re-solving the constraint equations at $t = 0$ makes little difference.

We carry out the entire evolution, before and after excision, in
the hyperbolic gauges. (The choice of $K_{\rm drive}$ has a
negligible effect on the evolution in this application.)  We
perform the same evolution on a $80^2$ grid, a $160^2$ grid, and a
$320^2$ grid.  On
the $320^2$ grid, a horizon appears in the pre-excision run
at $t = 44 M$, with instantaneous radius $r_{\rm AH} = 0.5 M$ and mass
$M_{\rm irr} = 0.77 M$, which are growing rapidly.  We excise at
time $t = 45.5 M$ and radius $r_{\rm ex} = 0.43 M$, so that
22\% of the total rest mass is still
outside the excision region, and 12\% is still outside the
apparent horizon (which now has radius $r_{\rm AH} = 0.73 M$). 
This matter quickly falls into the hole, and,
after evolving for $6 M$ with excision, the exterior spacetime becomes
a vacuum.  In Figure~\ref{rot_fig}, we check the ability of our
code to conserve mass and angular momentum during this phase of
the evolution.  The mass is well conserved on all three grids, but
the angular momentum slowly decreases with
time. Increasing resolution reduces this loss of $J$.  The violations
of the constraint equations also converge to zero as resolution is
increased.  We can evolve stably
for $t\gg 100 M$, but the loss of angular momentum is too great past
this point for the evolution to be reliable unless the grid exceeds
$320^2$.

%No matter escapes through the outer boundary, and gravity waves
%cannot carry away angular momentum in axisymmetry, so there is
%no physical explanation for the observed $J$ loss; it must represent
%numerical error.  We could find no significant improvement in the behavior
%of $J$ by moving the outer boundaries outward, evolving in a rotating
%coordinate frame, or evolving with several other standard gauge conditions. 
Figure~\ref{rot_fig} suggests that the angular momentum loss can be
controlled by increasing resolution. 
Moreover, we have already shown that our code can
conserve $J$ for an arbitrarily long time while evolving a Kerr black hole
in Kerr-Schild coordinates (see Fig.~\ref{ks_fig} and~\cite{ybs02}).  Given
this fact, we could eliminate the $J$ loss by transforming to
Kerr-Schild coordinates when we introduce excision.  Alternatively, we might
carry out the entire evolution in Kerr-Schild-like coordinates. 
(This would require developing gauge conditions which would force the
coordinate-system to maintain its Kerr-Schild-like character as the system
evolves.)  We are currently investigating these possibilities. 
In the meantime, we can already evolve such matter-black hole systems
long enough to tackle several interesting problems.

\section{Application:  The Collapse of Rapidly Rotating Stars}
\label{application}

\begin{table}
\caption{Equilibrium Star Configurations \\
 \ \ \ \ \ \ \ \ \ ($n = 1$, $M_0 = 0.2$).}
\begin{tabular}{ccccccccc}
\hline \hline  
 Star & $M^a$ &
 $\left.\right.\ R_{\rm eq}{}^b\ \left.\right.$ & $R_c{}^c$
 & $q{}^d$ & $T/|W|^e$ & $\Omega_c/\Omega_{\rm eq}^f$ & ${\cal R}{}^g$
 & \ \ Fate${}^h$\ \ \\ \hline  
\stara\ & 0.19 & 0.6 & 0.8 & 0.57 & 0.10 & 0.29 & 0.70 & BHND \\
\starb\ & 0.19 & 1.2 & 1.4 & 0.91 & 0.18 & 0.38 & 0.50 & BHND \\
\starc\ & 0.19 & 1.6 & 1.8 & 1.18 & 0.23 & 0.40 & 0.39 & NBH  \\
\hline \hline  
\label{star_table}
\end{tabular}

\raggedright

${}^a$ ADM mass 

${}^b$ coordinate equatorial radius

${}^c$ areal radius at the equator

${}^d$ $q = J/M^2$

${}^e$ ratio of kinetic to gravitational potential energy

${}^f$ ratio of central to equatorial angular velocity

${}^g$ ratio of polar to equatorial coordinate radius

${}^h$ BHND $=$ black hole, no disk; \ NBH $=$ no black hole

\end{table}

\begin{figure*}
\begin{center}
%\leavevmode \epsffile{fig9.ps}
\leavevmode
\epsfxsize=2.8in
\epsffile{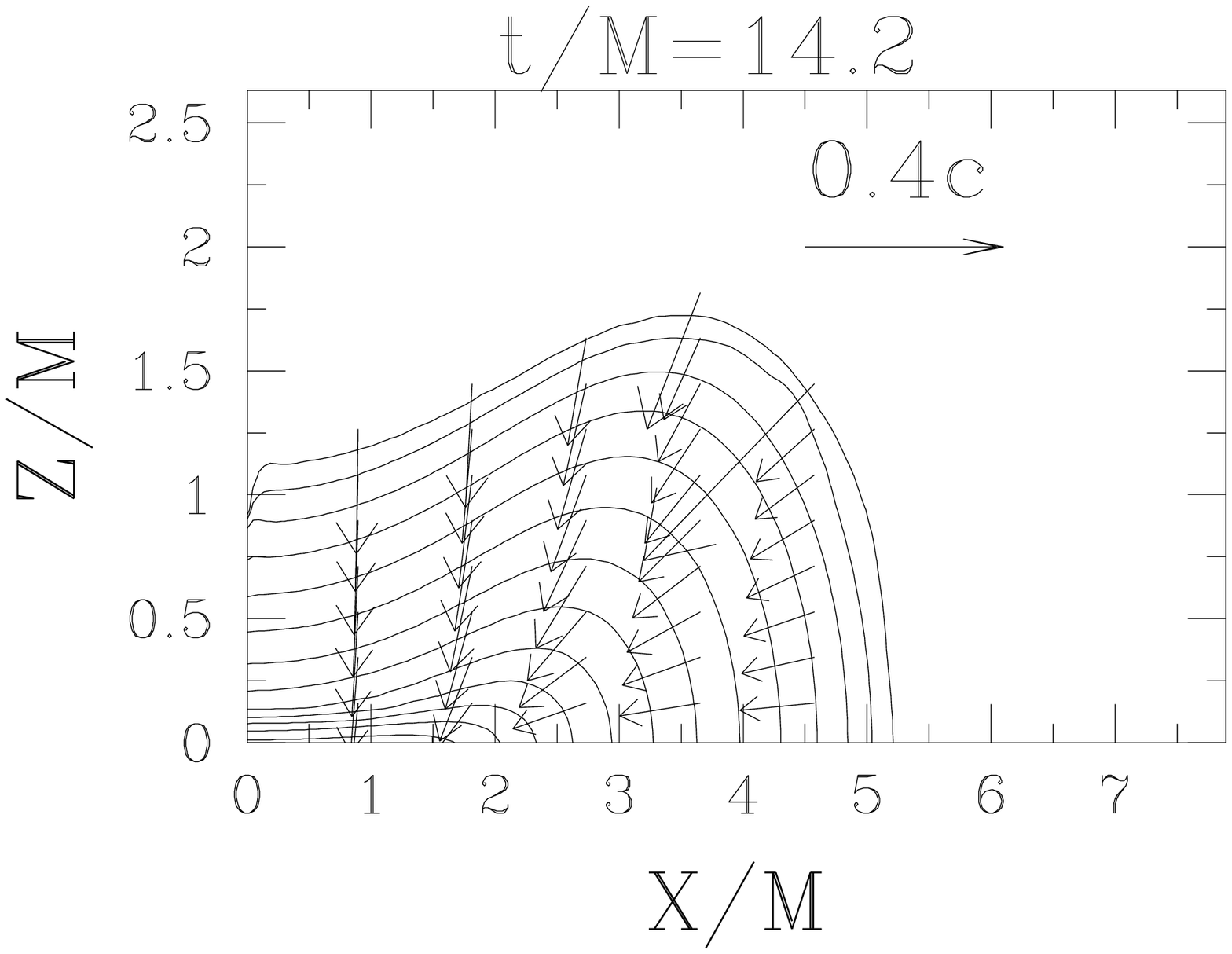}
\epsfxsize=2.8in
\leavevmode
\epsffile{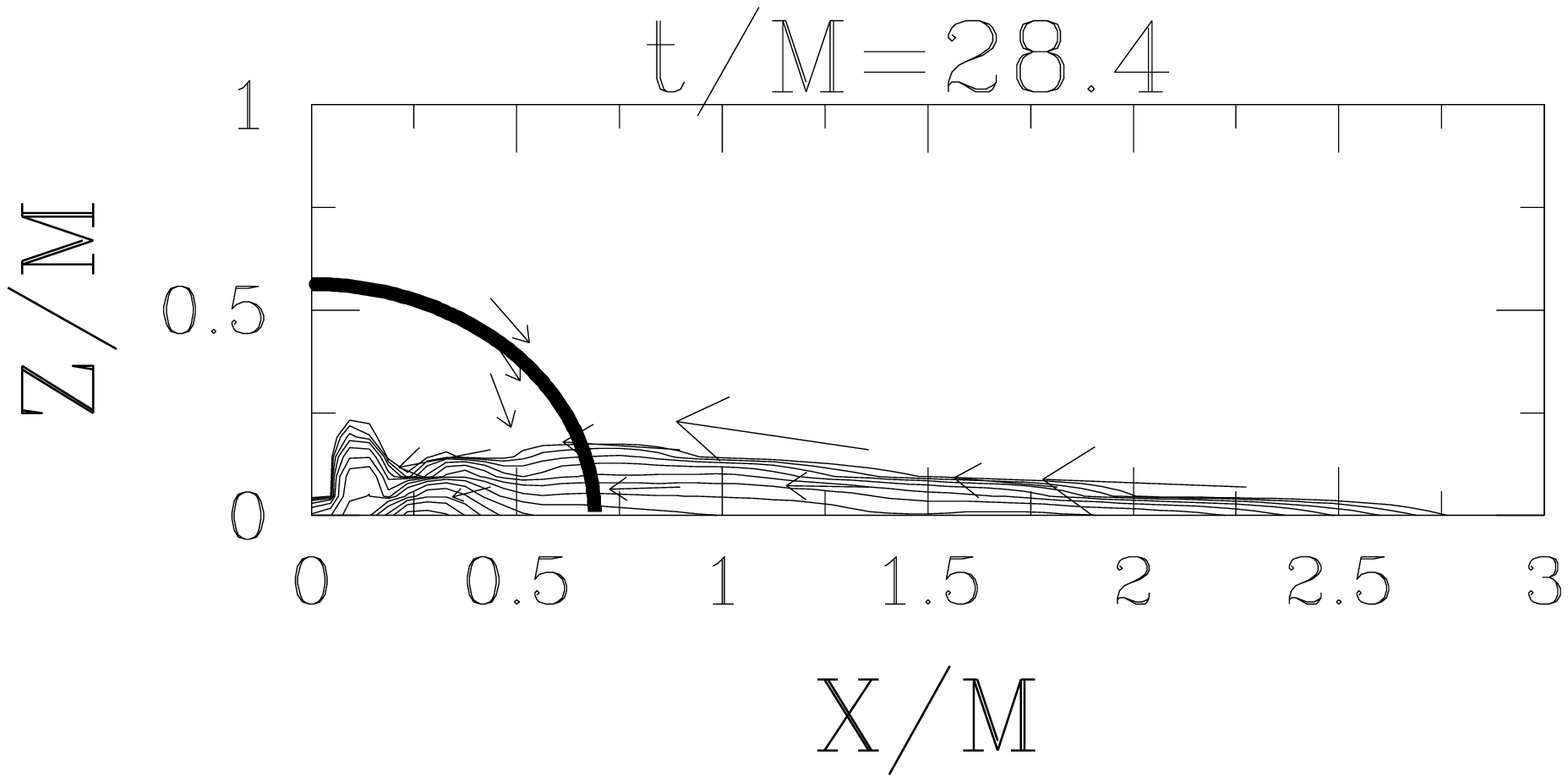}
\vskip20pt
\leavevmode
\epsfxsize=2.8in
\epsffile{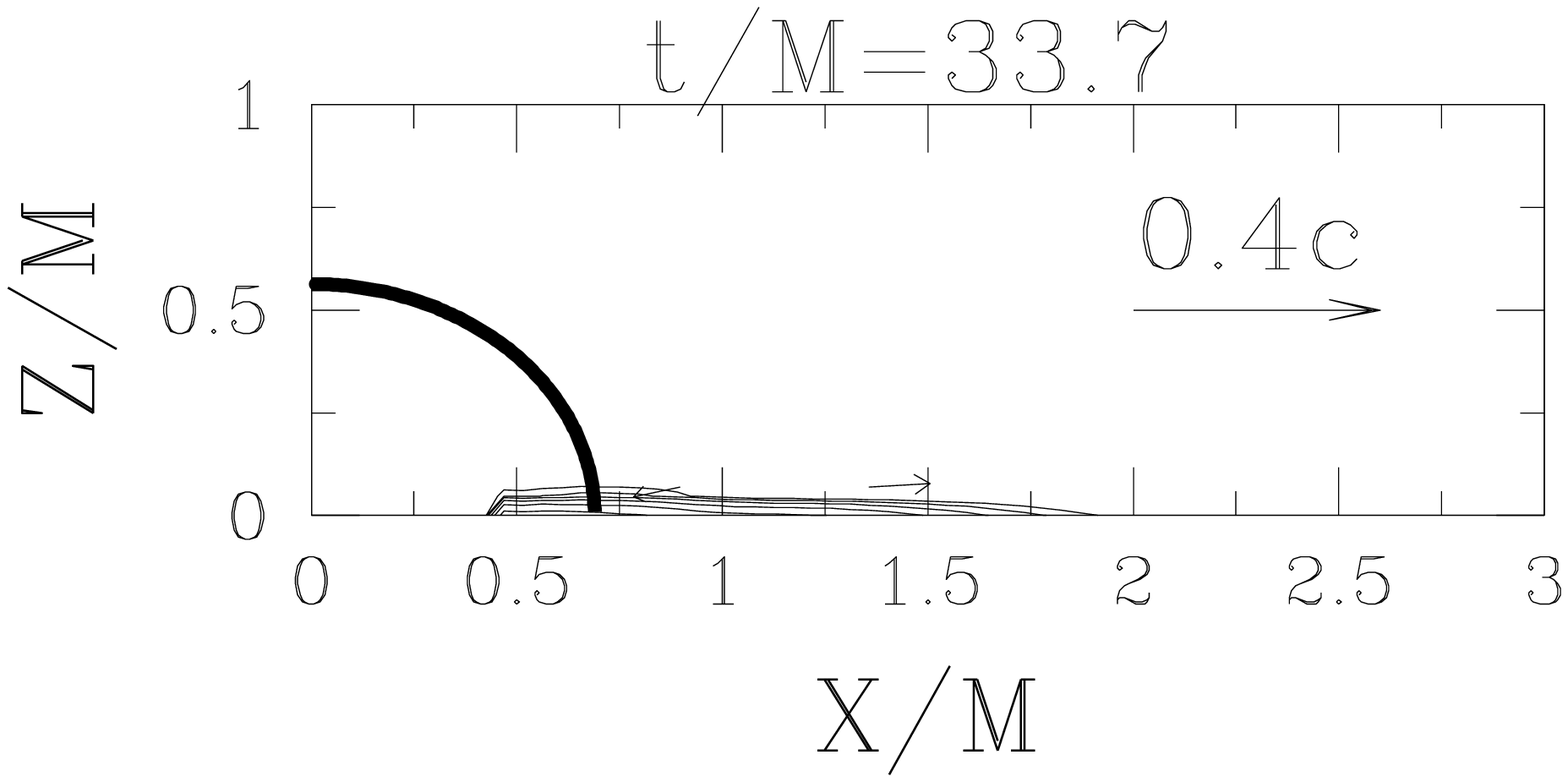}
\epsfxsize=2.8in
\leavevmode
\epsffile{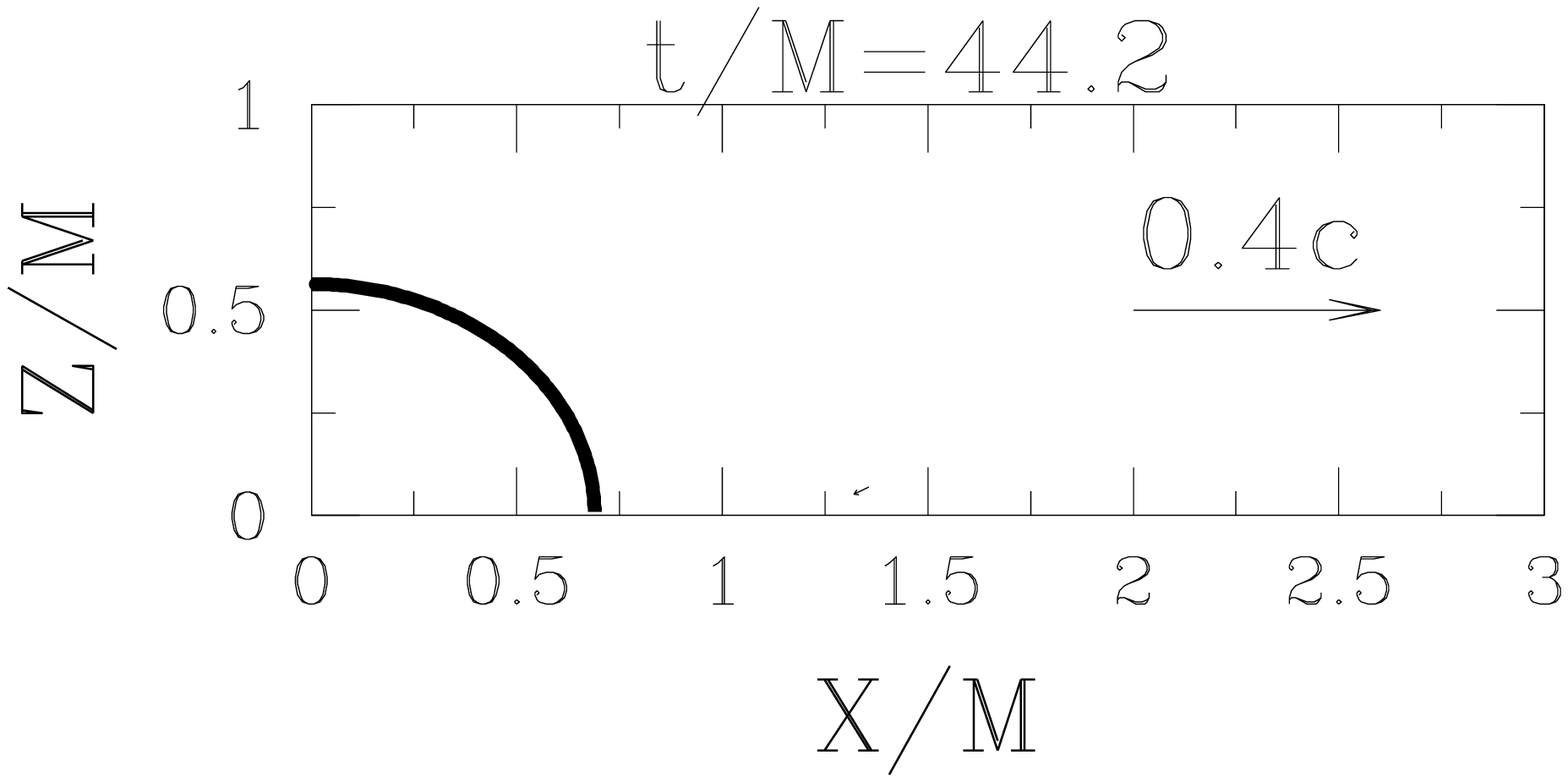}
\end{center}
\caption{ Snapshots of the rest-density contours and the
  velocity field $(v^x,v^z)$ in the meridional plane during
  the collapse of star \starb\ to a black hole.  The contour lines are
  drawn for $\rho_0=10^{-(0.2~j+0.1)} \rho_{0}^{Max} $ for $j=0,1,..,12$. 
  Prior to excision, $\rho_{0}^{Max}$
  is set equal to the instantaneous maximum value of $\rho_0$.  Afterwards,
  it is held at the maximum of $\rho_0$ at the time of excision. 
  Vectors indicate the local velocity field, $v^i$.  The thick curve in the
  last three frames marks the apparent horizon. On the last frame, the
  exterior spacetime is nearly a vacuum. }
\label{subk_fig1}
\end{figure*}

Tracking the collapse of rapidly rotating stars is one of the
most important applications of numerical general relativity. 
Such simulations determine the fate of collapse and provide a
test of the cosmic censorship conjecture~\cite{p69}.  If the star
collapses to a stationary black hole, the ``no-hair'' theorems
require that it settle down to a Kerr black hole.  In the
Kerr spacetime, the
singularity is covered by an event horizon only if $q\equiv J/M^2\le 1$;
otherwise the singularity is naked.  Rotating stars, on the other
hand, are not so restricted, and sufficiently rapidly rotating
stars will have $q > 1$.  When these stars collapse, it thus seems
conceivable that they could form naked singularities.  Alternatively,
if the cosmic censorship hypothesis~\cite{p69} is true, then the
collapse of the whole system
must somehow be averted.  This can happen if the star loses angular
momentum as it collapses, either by gravitational wave emission or
by shedding matter with high specific angular momentum, so that the
final black hole has $q < 1$.  A naked singularity can also be
averted if the collapse of a $q > 1$ star is always halted by
centrifugal forces, so there will be no black hole and no singularity
at all.  Nakamura~\cite{n02} has pointed out that
a centrifugal barrier could protect cosmic censorship in this way. 
Assuming no mass or angular momentum are shed during the collapse,
the radius $R_b$ at which the centrifugal force balances the
gravitational force will be
\begin{equation}
{M\over R_b^2} \sim {J^2\over M^2R_b^3}\ ,
\end{equation}
so that
\begin{equation}
R_b \sim M q^2\ .
\end{equation}
Nakamura argues that, if $q<1$ (i.e., the star is {\it subKerr}), the
star will already be inside a
black hole before rotation can halt the collapse.  For $q>1$
(i.e., the star is {\it supraKerr}), the
collapse will be halted at a radius larger than $M$, and no black hole forms.

Shapiro and Teukolsky~\cite{acst94} have studied the collapse in full
general relativity of axisymmetric tori consisting of collisionless matter,
and have found that black
holes form only from subKerr initial configurations. 
The first numerical simulations of the collapse of rotating relativistic
fluid stars were carried out in axisymmetry by Nakamura~\cite{n81} and
Nakamura and Sato~\cite{ns81}.  They found that a black hole
forms only when a subKerr star collapses.  (For stars with $q$ within
5\% of the critical
value, Nakamura~\cite{n81} could not determine the final fate and could
not exclude the possibility of a naked singularity.)  Stark and Piran
\cite{sp85} also performed simulations which showed $q\!\sim\! 1$ to
be the critical point of demarcation between collapse and bounce. 
Shibata~\cite{s00} performed a detailed study of the collapse and
bounce of subKerr stars in axisymmetry.  These hydrodynamic studies
did not (and
sometimes could not) study in detail the fate of the matter in the outer
layers of the star when a black hole forms.  More recently,
Shibata~\cite{s03} has studied the collapse to black holes of uniformly rotating
polytropes spinning at the mass-shedding limit.  He finds that, for
polytropic indicies $2/3<n<2$, the star collapses to a Kerr black hole
with no appreciable disk. 
By using high resolution, he is able to follow the system for
$\Delta t\sim 20 M$
after an apparent horizon is first located.  This time approaches the
limit of reliable evolution without excision, but in this case
it is long enough to see all the matter fall into the hole.  By contrast,
Shibata and Shapiro~\cite{ss02_1} considered the collapse of an $n = 3$
polytrope spinning uniformly at the mass-shedding limit.  Such a
configuration is nearly Newtonian ($R_{\rm eq} = 620 M$) at the onset
of collapse, and it forms an appreciable disk ($M_{\rm D}/M \approx 0.1$)
around the final black hole.  While the final disk mass can be
estimated from the angular momentum distribution of the outermost regions
(see also~\cite{ss02_2}), and also by extrapolating the growth of the
black hole horizon to late times, it is not possible to follow the
final relaxation to a stationary state without excision or to probe for
nonaxisymmetric instabilities that may arise in the ambient disk~\cite{sms1}

Our excision code should be well suited to finding the final state of any
rapidly rotating stellar collapse---not only for determining whether
or not a black hole forms, but also for determining how much rest
mass escapes collapse if one does form.  To explore this capability,
we take differentially
rotating polytropes as our initial data, so that we can study both subKerr
and supraKerr cases.  Differential rotation is naturally produced in
supernova core collapse~\cite{zm97}, accretion induced collapse of
white dwarfs to neutron stars~\cite{ll01}, and binary neutron star
coalescence~\cite{rs92,su00,fr02}.  Our adopted rotation law is
\begin{equation}
  u^tu_{\phi} = R_{\rm eq}^2 A^2(\Omega_c - \Omega)\ ,
  \label{rot_law}
\end{equation}
where $\Omega$ is the angular velocity of the fluid, $\Omega_c$ is the
value of $\Omega$ on the rotation axis, $R_{\rm eq}$ is the equatorial
coordinate radius.  The parameter $A$ measures the degree of differential
rotation and is chosen to be unity for all
cases below, so that the centers of our stars rotate
about three times faster than their equators.  We take the $z$-axis
to be the rotation axis, and define the cylindrical coordinate radius
$\varpi = \sqrt{x^2 + y^2}$. 
In the Newtonian limit, Eq.~(\ref{rot_law}) reduces to the so-called
``$j$-constant'' law~\cite{eriguchi85}
\begin{equation}
  \Omega = {\Omega_c\over 1 + {\varpi^2\over R_{\rm eq}^2 A^2}}\ .
\end{equation}
We choose a polytropic index $n = 1$, and take our
initial stars to be sufficiently compact so that the collapse does
not span a large dynamic range.  Accordingly, we are able to use a single,
modest grid for each run.  As in~\cite{sp85,s00,dmsb03}, we induce
collapse by depleting the initial pressure by a factor: 
$P\rightarrow f_P P$.  Below, we show results for $f_P = 0.01$. 
While this form of artificially-induced collapse does not correspond
to any realistic astrophysical scenario, there
are several situations in which an ``effective''
pressure depletion does occur.  For example, the collapse of the
core of a massive star which produces a supernova is brought about
by the removal of pressure support both from photo-dissociation of
iron-nickel
nuclei and the neutronization of the core (de-leptonization). 
Phase transitions in neutron stars,
such as a transition to quark matter, or rapid de-leptonization via
neutrino cooling, could also have the effect of
inducing pressure depletion.  We choose $f_P = 0.01$ to make pressure
forces unimportant in comparison with centrifugal forces and gravity.
After depleting pressure from the star, we re-solve the constraint
equations to produce valid initial data.  This process of depleting
pressure and re-solving the constraints causes $M$ and $J$ to drop
by a few percent, while $J/M^2$ changes by one percent or less.

Table I lists the equilibrium stars used to construct our initial data. 
These initial data were generated using the code of~\cite{cst92}. 
Each star has
the same rest mass $M_0 = 0.2$, so our stars are members of a
sequence uniquely defined by $n = 1$, $A = 1$, $M_0 = 0.2$.  This
sequence crosses $q = 1$ at one point, between our second and
third stars, stars \starb\ and \starc.  We expect to find a qualitative
difference in the behavior of stars \starb\ and \starc.

Star \stara\ is exactly the star studied in the previous section.  It is
dynamically unstable and collapses without pressure depletion to a
Kerr black hole with no disk.  Not surprisingly, this is
also found to be the behavior when pressure is depleted.  We will
concentrate below on stars \starb\ and \starc.  We begin with simulations
in axisymmetry and then discuss simulations in full three dimensions.

\begin{figure}
\epsfxsize=2.8in
\begin{center}
\leavevmode \epsffile{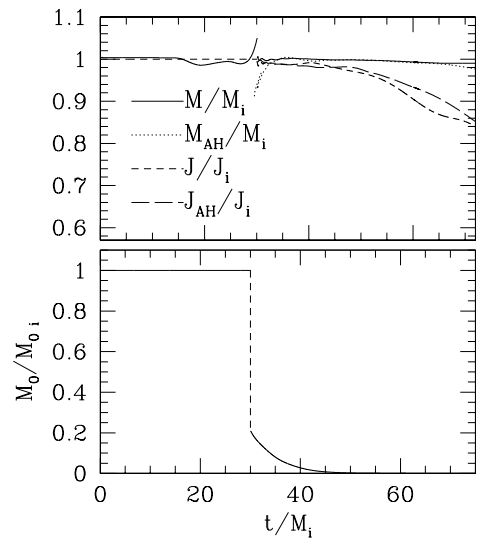}
\end{center}
\caption{ Diagnostics for the collapse of star \starb.  Above,
  we show the evolution of $M$ and $J$ calculated from integrations
  of the exterior spacetime and from measurements of the geometry of
  the horizon.  Below, we plot the total rest mass on the grid,
  normalized to its initial value.  Rest mass is conserved prior
  to excision.  At $t = 30 M$, we excise a region from the middle
  of the grid.  This cuts out the matter inside this region, which
  accounts for 80\% of the total rest mass.  Over the next $20 M$,
  the remaining rest mass falls into the excision zone, leaving a
  vacuum being evolved in the outside region. }
\label{subk_fig2}
\end{figure}
\begin{figure*}
\begin{center}
%\leavevmode \epsffile{fig11.ps}
\epsfxsize=2.8in
\leavevmode
\epsffile{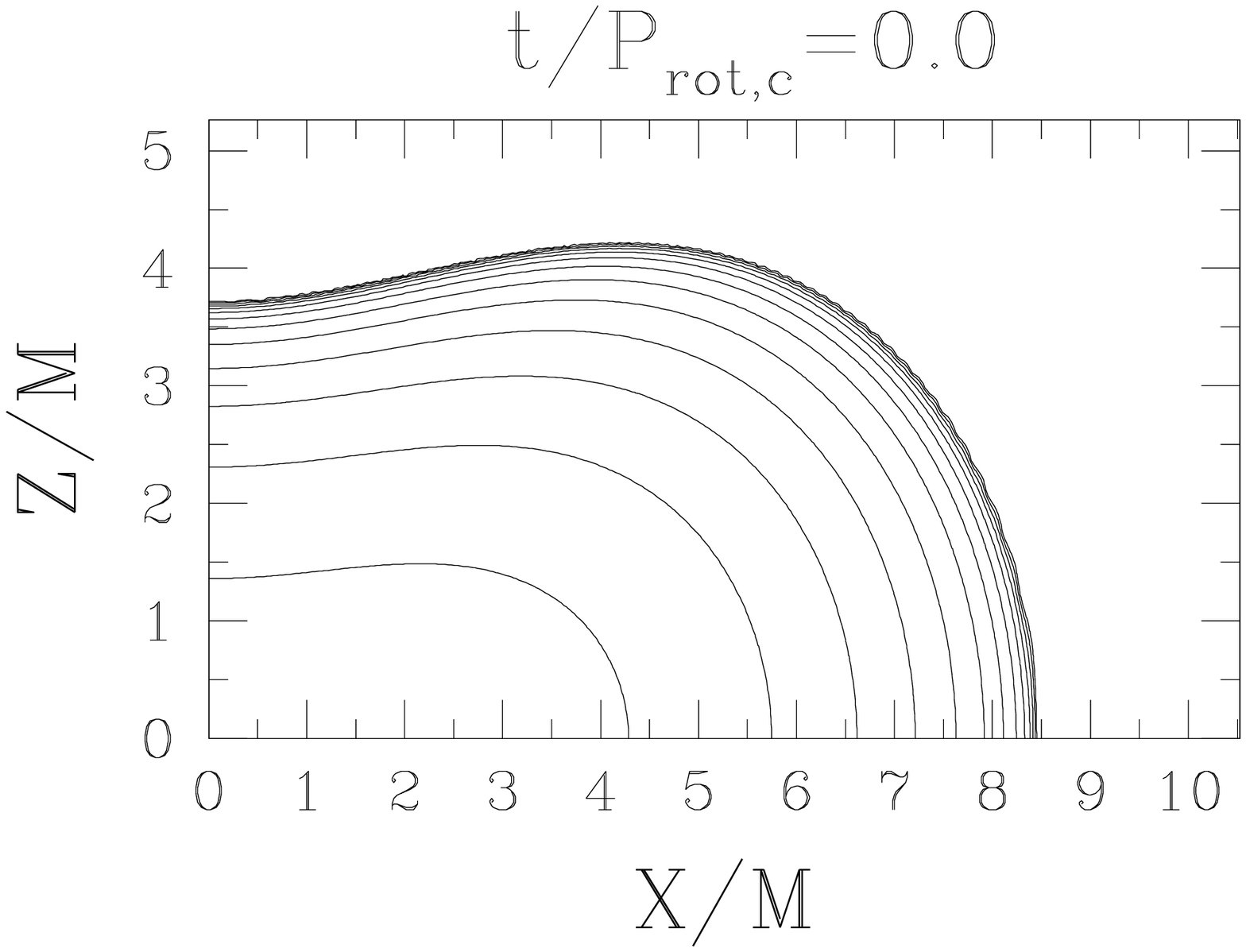}
\epsfxsize=2.8in
\leavevmode
\epsffile{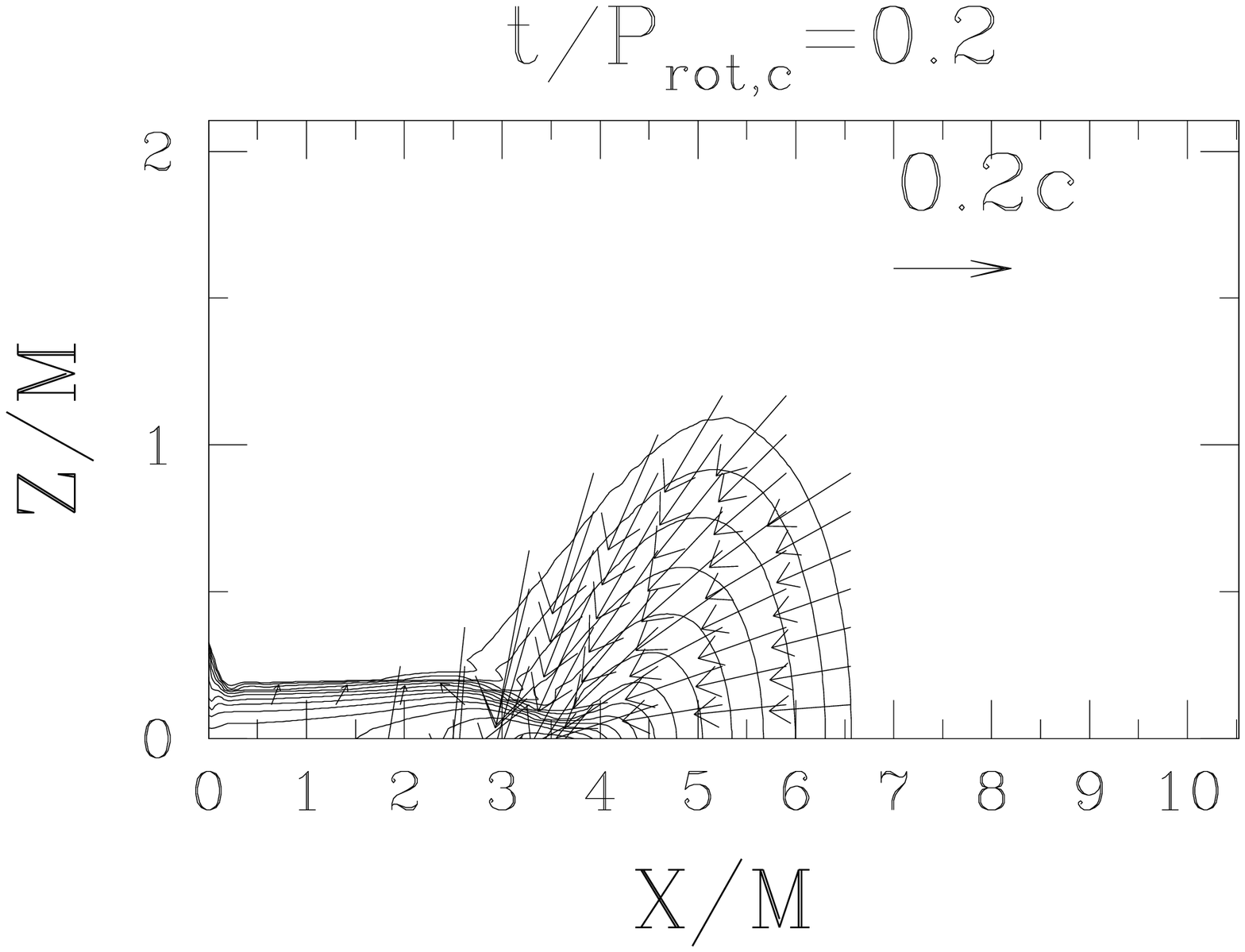}

\vskip10pt

\leavevmode
\epsfxsize=2.8in
\epsffile{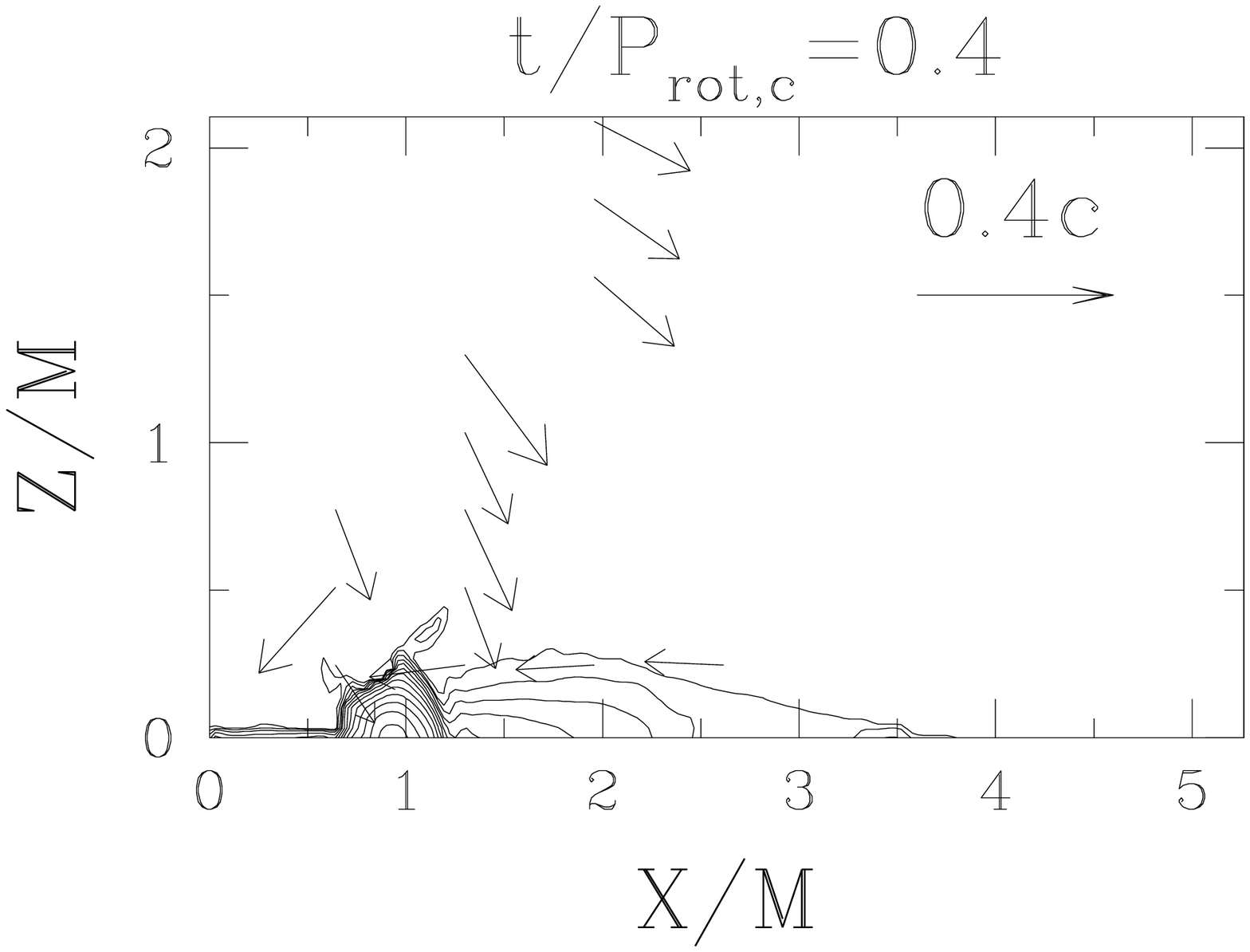}
\epsfxsize=2.8in
\leavevmode
\epsffile{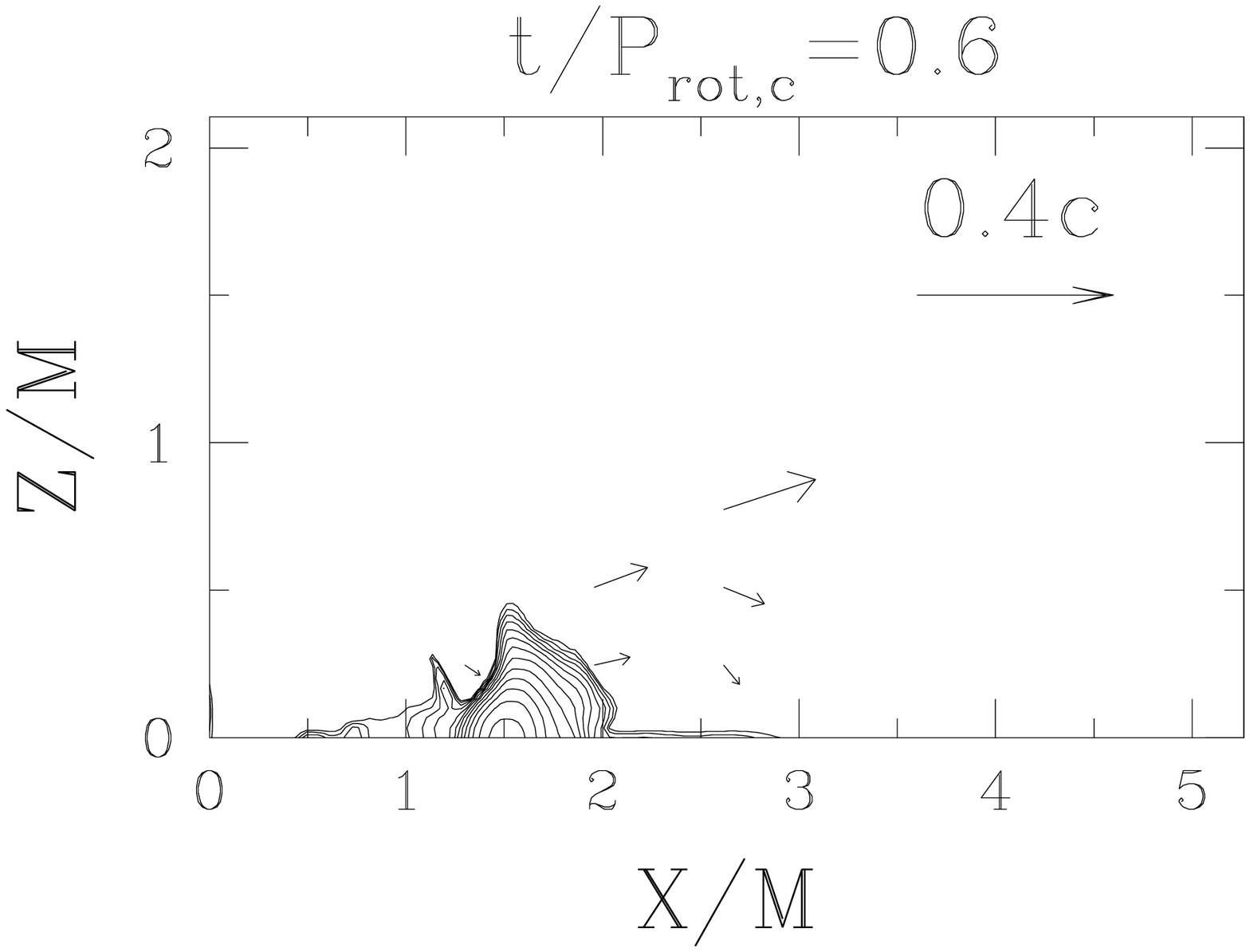}

\vskip10pt

\leavevmode
\epsfxsize=2.8in
\epsffile{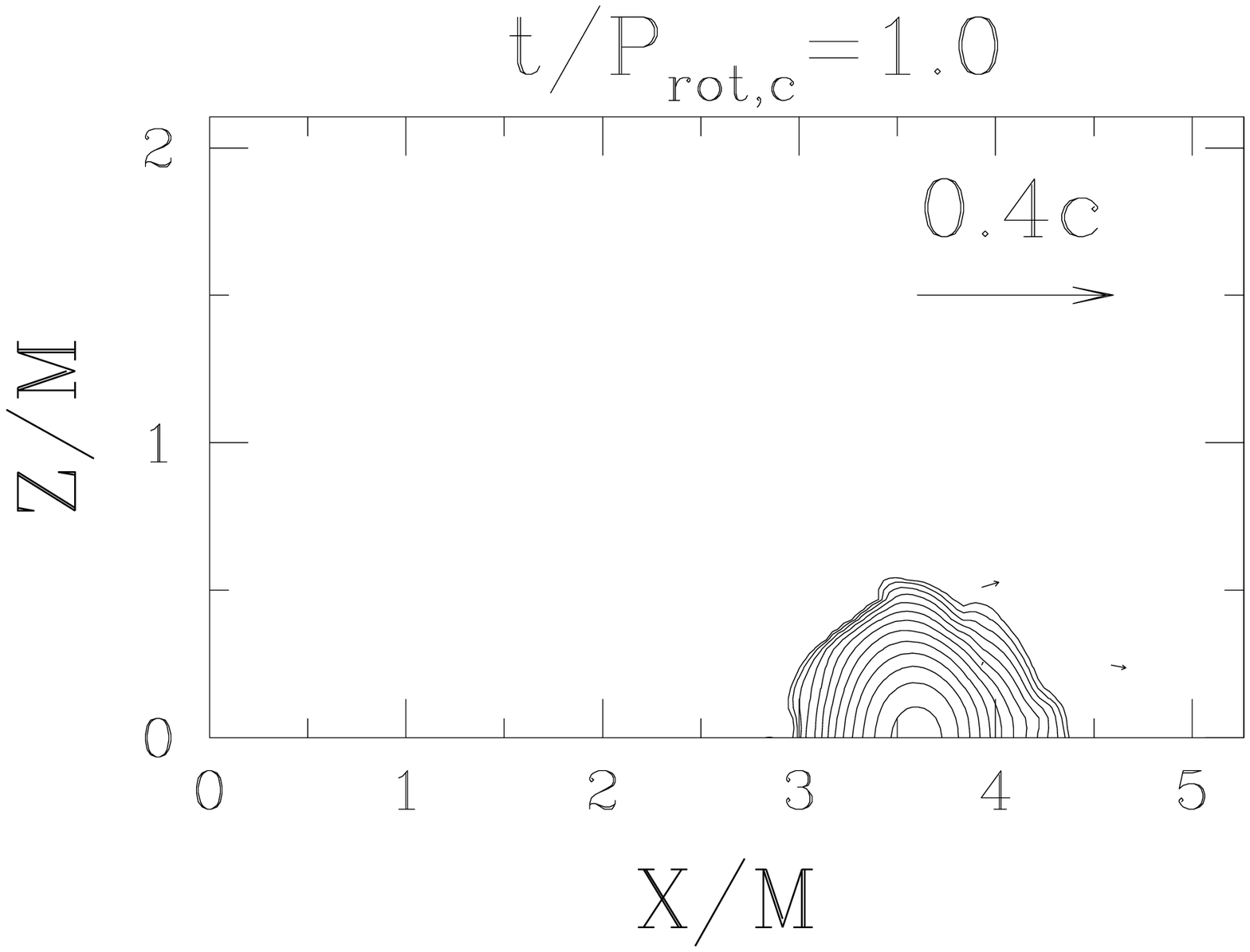}
\epsfxsize=2.8in
\leavevmode
\epsffile{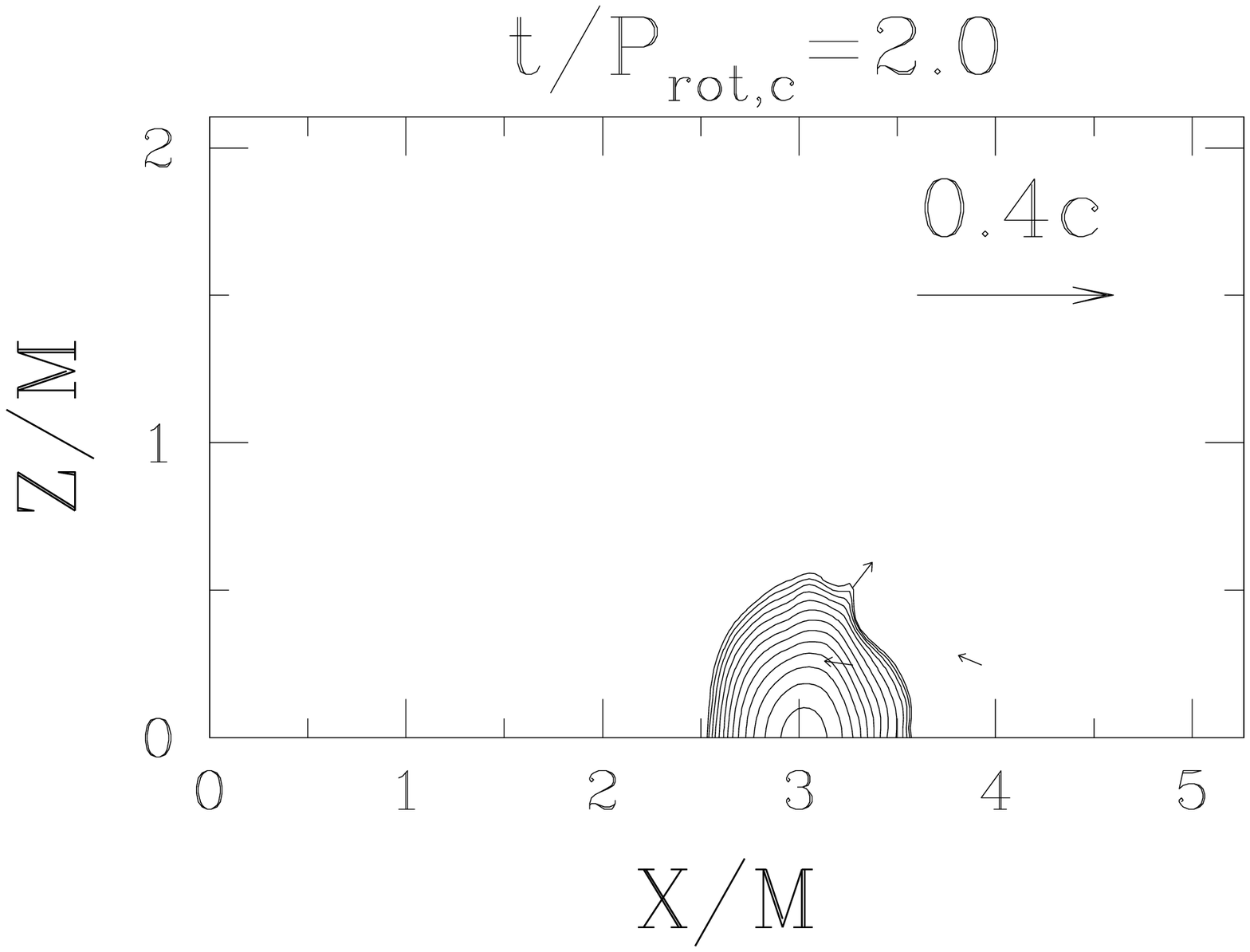}
\end{center}
\caption{ Snapshots of the rest density contours and the
  velocity field $(v^x,v^z)$ in the meridional plane during the axisymmetric
  collapse of star \starc\ to a torus. The contours are set
  as in Figure~\ref{subk_fig1}. 
  Some velocity arrows appear outside the contours
  because the density there is very small but nonzero.
  Time is normalized to the initial central rotation period of the
  star, $P_{\rm rot,c} = 98 M$. }
\label{supk_fig1}
\end{figure*}

\subsection{SubKerr Collapse}

\begin{figure*}
\epsfxsize=5.2in
\begin{center}
\leavevmode \epsffile{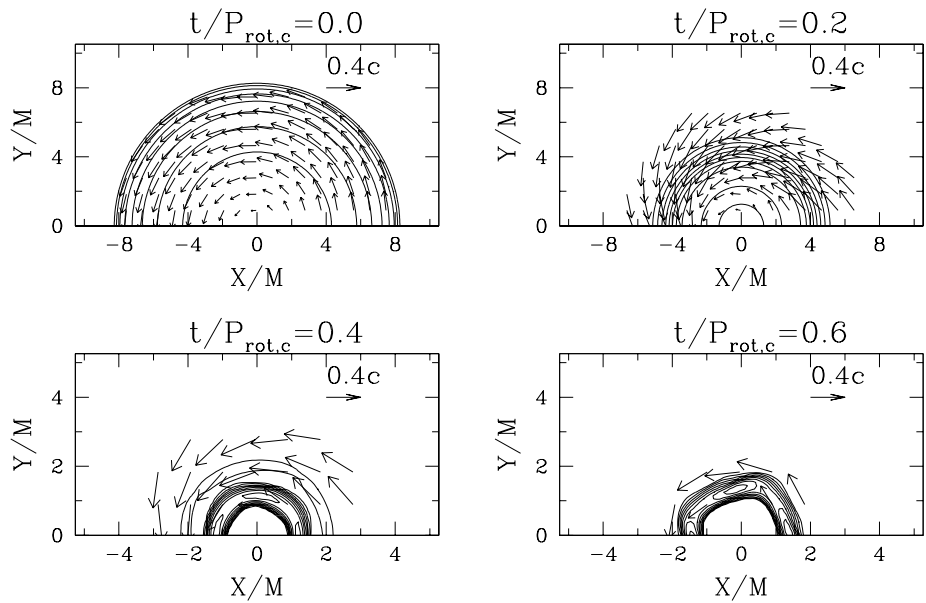}
%\epsfxsize=2.5in
%\leavevmode
%\epsffile{fig12a.ps}
%\epsfxsize=2.5in
%\leavevmode
%\epsffile{fig12b.ps}
%\leavevmode

%\vskip10pt

%\epsfxsize=2.5in
%\epsffile{fig12c.ps}
%\leavevmode
%\epsfxsize=2.5in
%\epsffile{fig12d.ps}
\end{center}
\caption{ Snapshots of the rest-density contour lines for $\rho_0$ and the
  velocity field $(v^x,v^y)$ in the equatorial plane for the 3D
  collapse, bounce, and fragmentation of star \starc.  The contours and
  time normalization are set using the same rule as in Figures~\ref{subk_fig1}
  and~\ref{supk_fig1}. Note that the origin of the system is now shifted to
  the middle of the $x$-axis in this plot. }
\label{supk_fig3}
\end{figure*}

\begin{figure}
\epsfxsize=2.8in
\begin{center}
\leavevmode \epsffile{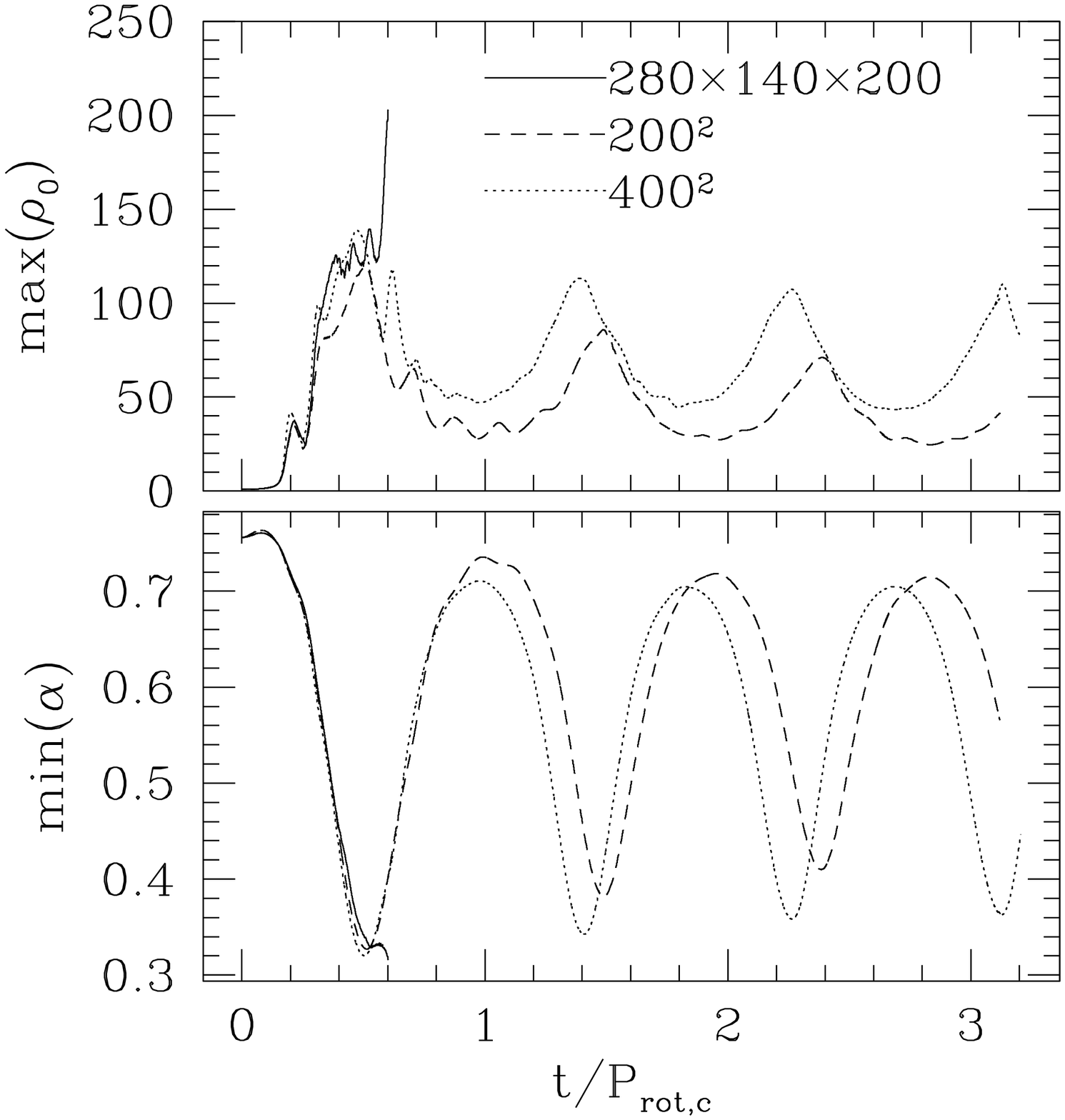}
\end{center}
\caption{ The maximum value of $\rho_0$ and the minimum value of
  $\alpha$ during the evolutions of star \starc\ on different grids, plotted
  as a function of the initial central rotation period $P_{\rm rot,c}$.
  The two 2D runs are qualitatively similar.  The 3D run behaves
  similarly to the 2D runs for about the first $0.5 P_{\rm rot,c}$.
  Thereafter a nonaxisymmetric instability develops, and the collapsed
  star fragments. }
\label{supk_fig2}
\end{figure}
Star \starb\ has $J/M^2 = 0.9$, so it is subKerr.  Its collapse
in axisymmetry is
shown in Fig.~\ref{subk_fig1}.  We evolve on a $300^2$ grid
with outer boundaries at $14 M$.  At $t = 28.4 M$, we
locate an apparent horizon with $r_{\rm AH} = 0.62 M$, $M_{\rm irr} = 0.72 M$. 
We excise at $t = 29 M$, at which time $M_{\rm irr} = 0.74 M$, 22\%
of the rest mass is outside our excision zone, and 15\% is outside the
apparent horizon.  The horizon circumferences at this time are
in the ratio $C_{\rm pol}/C_{\rm eq} = 0.76$, which,
if this were a stationary Kerr horizon, would correspond to
$q = 0.92$.  We continue evolving with an excision boundary at
radius $r_{\rm ex} = 0.08$.  All of the matter falls into the hole
within $20 M$ after excision is introduced.  We evolve for
an additional $20 M$ after this.  We find no signs of numerical
instability.  Mass conservation is excellent (the amount lost due
to gravitational radiation is below 0.1\%), but
the gradual loss of angular momentum noted in Section~\ref{rot_collapse} is
present, as can be seen in Figure~\ref{subk_fig2}.  We stop evolving
when the total
angular momentum drops below 80\% of its initial value.  The
final state of the system has, however, been entirely determined
well before this time.

\subsection{SupraKerr Collapse}

Star \starc\ has $J/M^2 = 1.2$.  We remove the star's pressure support
and evolve.  In Figure~\ref{supk_fig1}, we show the results of
a $400^2$ axisymmetric run with boundaries at 13 M. 
With its pressure support removed, the star immediately flattens along
the $z$-axis and moves inward in $\varpi$.  This inward motion toward
the axis is halted by centrifugal forces.  As seen in the
upper right panel of Fig.~\ref{supk_fig1}, the inner region of the
star stops collapsing before the outer region, so a strong shock is
formed.  The star then expands into a torus whose radius oscillates
with a period close to the initial central rotation period.  We show
the effects of this oscillation on the maximum rest density and the
minimum lapse in Fig.~\ref{supk_fig2}.  We follow the torus for
three oscillations during which time all our constraints are satisfied
to better than 10\%.  The angular momentum $J$ is conserved identically
by our no-excision
axisymmetric code, but we do find that the ADM mass $M$ decreases
gradually with time.  This decrease cannot be accounted for by
the small flux of rest mass and gravity waves out of the computational
domain; the loss therefore represents numerical error.  We stop our evolution
after three oscillation periods because $M$ has decreased by $\sim$15\%. 
To check that the evolution is qualitatively correct, we
%include the results for a simulation of the same star on a $200^2$ grid in
%Fig.~\ref{supk_fig2}.
performed the same run on a $200^2$ grid and found that the collapse,
torus-formation, and oscillation of the star are very similar at this
resolution.

The torus formed in the above simulation could be subject to various
non-axisymmetric instabilities.  If the rotating torus fragments, the system
may produce a large gravitational wave signal
(``splash radiation''~\cite{rrw74}). 
It is therefore necessary to perform the above simulation in
3+1 dimensions.  We perform this simulation using a
$280\times 140\times 200$ grid, with boundaries at
$[-13 M,13 M]\times[0,13 M]^2$, where we use equatorial and $\pi$-symmetry. 
The results are shown in Fig.~\ref{supk_fig3}.  The
collapse, flattening, and formation of the torus occur as in the 2D
runs.  Then the torus quickly fragments into four clumps
symmetrically located about the origin, roughly $90^{\circ}$ apart.  As these
clumps collapse, they ultimately become too small to be evolved accurately on
our grid.  We conserve $M$ and $J$ to better than 10\% throughout
the integration shown, and we terminate the calculation when our errors
exceed these bounds.  To check this result, we have performed the
same run on $140\times 70\times 180$ and $100\times 50\times 100$
grids.  In each case, the torus fragments into four pieces
$90^{\circ}$ apart.  In
Fig.~\ref{supk_fig2}, we compare the behavior of the maximum of
$\rho_0$ and the minimum of $\alpha$ for the evolution of star \starc\
in 3D to their behavior in 2D on $200^2$ and $400^2$ grids.

It has been pointed out by Truelove {\it et al.} \cite{tkmhhg97} that
spurious fragmentation may occur in a numerical simulation if
the Jeans length is not well resolved.  The Jeans length is given
by
\begin{equation}
\lambda_J \sim \sqrt{{\pi c_s^2\over \rho}}\ ,
\end{equation}
where $\rho$ is the density (Mass-energy density and rest-mass density
are nearly equal.) and $c_s = \sqrt{{dP\over d\rho}}$
is the sound speed.  We can get a lower bound on $\lambda_J$ by
ignoring the large amount of shock heating, which increases $c_s$,
and considering adiabatic compression.  Accordingly, for an $n = 1$,
$\Gamma = 2$ fluid, $P = \kappa\rho_0^2$, where $\kappa = 0.01$ due
to our pressure depletion.  Fragmentation occurs when
$\rho\approx\rho_0\approx 3$, so $\lambda_J\sim 0.25 = 15\Delta X$. 
(Shock heating increases this coefficient.) 
Our resolution is then quite sufficient to resolve the Jeans length.

\begin{figure}
\epsfxsize=2.8in
\begin{center}
\leavevmode \epsffile{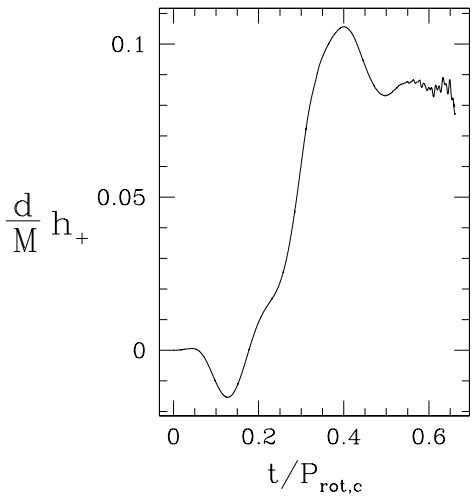}
\end{center}
\caption{ The gravitational wave amplitude $h_+$, at a distance $d$ from
  the source, for the 3D collapse,
  bounce, and fragmentation of star \starc.  We compute $h_+$ at the
  point (11.6 $M$, 0, 0). }
\label{gwaves}
\end{figure}

We could not determine the final fate of this system.  The four clumps
may continue to collapse to black holes, or this collapse may be halted
by heating-induced pressure.  The system will certainly emit substantial
amounts of gravity waves, both during the bounce and oscillation of the
initially axisymmetric torus and during its rotation following fragmentation. 
To see this, we measured the gauge invariant
Moncrief variables $\psi_{lm}$ (or Zerilli functions) at the outer
part of the grid~\cite{m74}.  We also measure the amplitudes of the
two gravitational wave polarizations $h_+$ and $h_{\times}$ on the
$x$-axis at the edge of our grid.  Since the outer part of the grid is not
in the wave zone, our measurements are only approximate. 
We find that the dominant mode of the emission is $l = 2$, $m = 0$, the
quadrupole radiation generated by the axisymmetric collapse and bounce
of the torus.  The second largest modes, which are an order of magnitude
smaller than the dominant mode, are $l = 4$, $m = 0$ (octopole radiation
from the axisymmetric collapse) and $l = 4$, $m = \pm 4$ (octopole
radiation generated by the rotation of the four clumps).  In
Fig.~\ref{gwaves}, we plot $h_+$ on the $x$-axis, which contains
contributions from all modes. 
%We find that
%gravitational wave emission is particularly strong in the $m = 0$ and $m = 4$
%modes.  The $m = 0$ waves are quadrupole ($l = 2$) dominated, and are
%generated
%by the collapse and bounce of the torus.  The $m = 4$ waves are octopole
%($l = 4$) dominated and are generated by the rotation of the four clumps. 
The observed amplitude of this radiation from a star at a
distance $d$ from the Earth would be
\begin{equation}
  h \sim 10^{-22} \left({M\over M_{\odot}}\right)
                  \left({d\over 100 {\rm Mpc}}\right)^{-1}
\label{gwave_amp}
\end{equation}
The final evolution
of this very interesting system can only be undertaken using a
finer grid, presumably by employing adaptive mesh refinement (AMR),
and an improved shock-handling scheme in our code.

\section{Discussion and Conclusions}
\label{discussion}

We have constructed a code to study the collapse of astrophysical
objects to black holes by evolving the full coupled Einstein-hydrodynamics
system in both 2+1 (axisymmetry) and 3+1 dimensions.  When a black hole
appears, it is treated by introducing an excision boundary well inside
the horizon.  Our code is stable and convergent for all of the test
problems and applications presented here.  As a test application,
we study the collapse of rapidly rotating stars.  Our conclusions
regarding their ultimate fate agree
with those of Nakamura~\cite{n81} and of Stark and Piran~\cite{sp85}---
namely, that spinning stars deprived of their pressure support will
collapse directly to black holes only if they are subKerr.  This is
the same behavior observed for spinning configurations of collisionless
matter~\cite{acst94}.  We also were able to study
the final state of the subKerr collapses by using our excision algorithm
to extend the evolution far beyond what could be achieved without it. 
We find that even for a rapidly rotating star with $q = 0.9$, all the rest mass
falls immediately into the hole, with no disk formation, in agreement with
Shibata~\cite{s03}. 
For the case of supraKerr collapse, we found
that the collapsing star hits a centrifugal barrier and bounces,
forming a torus which fragments due to a nonaxisymmetric instability
into four pieces.  With our current computational resources, we were
unable to determine the final fate of the four clumps.  Systems like this
one are sufficiently interesting as gravitational wave
sources, that they should be pursued by further investigation with finer
resolution, including AMR.

Considering the stability of our excision algorithm over such a variety of
applications,
we believe that it has great promise as a tool for relativistic astrophysics
involving the simultaneous presence of hydrodynamic matter and black holes. 
Our current post-excision algorithm exhibits a gradual spurious decrease
in total angular momentum when applied at moderate resolution. 
However, this problem is not present
in all coordinate systems (e.g. Kerr-Schild) and is reduced as the resolution
is increased.  We are currently investigating a number of ways to
improve our algorithm and to apply it to other 2D and 3D problems of
astrophysical interest.

\acknowledgments

It is a pleasure to thank Charles Gammie, Yuk-Tung Liu, and Branson Stephens
for useful discussions.  Most of the calculations
were performed at the National Center for Supercomputing Applications at the
University of Illinois at Urbana-Champaign (UIUC).  The remaining calculations
were performed at National Center for High-performance Computing in Taiwan.
This paper was supported in part by NSF Grants PHY-0090310 and PHY-0205155
and NASA Grant NAG 5-10781.

%\begin{references}

\end{document}